\documentclass{aa}  

\usepackage{appendix}

\usepackage{graphicx}
\usepackage{txfonts}
\usepackage{siunitx}
\usepackage{natbib}

\usepackage{amsmath}
\usepackage{mathtools}

\usepackage{tikz}
\usepackage{verbatim}
\usepackage{microtype}
\usepackage{adjustbox}

\usepackage{subcaption}
\usepackage[colorlinks,citecolor=blue, linkcolor=blue, urlcolor=blue,bookmarks=false,hypertexnames=true]{hyperref}

\usetikzlibrary{shapes,arrows,positioning,calc}

\bibpunct{(}{)}{;}{a}{}{,} 

\graphicspath{{./}{figures/}}

\usepackage{subfiles}

\begin{document} 

\title{Active Galactic Nuclei catalog from the AKARI NEP Wide field}

  \author{
         Artem Poliszczuk \inst{1} \fnmsep\thanks{\email{\url{artem.poliszczuk@ncbj.gov.pl}} 
         }
        \and
        Agnieszka Pollo \inst{1,2}
        \and
         Katarzyna Małek \inst{1,3}
        \and
        Anna Durkalec \inst{1}
        \and
        William J. Pearson \inst{1}
        \and
        Tomotsugu Goto \inst{4}
        \and
        Seong Jin Kim \inst{4}
        \and
        Matthew Malkan \inst{5}
        \and
        Nagisa Oi \inst{6}
        \and
        Simon C.-C. Ho \inst{4}
        \and
        Hyunjin Shim \inst{7}
        \and
        Chris Pearson \inst{8,9,10}
        \and
        Ho Seong Hwang \inst{11}
        \and
        Yoshiki Toba \inst{12, 13, 14}
        \and
        Eunbin Kim \inst{15}
        }

  \institute{
  National Centre for Nuclear Research, Pasteura 7, 02-093 Warsaw, Poland
  \and
  Astronomical Observatory of the Jagiellonian University, ul.Orla 171, 30-244 Krakow, Poland
  \and
   Aix Marseille Univ. CNRS, CNES, LAM, Marseille, France
  \and
  Institute of Astronomy, National Tsing Hua University, 101, Section 2. Kuang-Fu Road, Hsinchu, 30013, Taiwan
  \and
  Department of physics and astronomy, UCLA, 475 Portola Plaza, L.A., CA 90095-1547, USA
  \and
  Tokyo University of Science, 1-3, Kagurazaka Shinjuku Tokyo 162-8601 Japan
  \and
  Department of Earth Science Education, Kyungpook National University, Daegu 41566, Korea
  \and
  RAL Space, Rutherford Appleton Laboratory, Chilton, Didcot, Oxfordshire OX11 0QX, UK
  \and
  School of Physical Sciences, The Open University, Milton Keynes, MK7 6AA, UK
  \and
  Oxford Astrophysics, University of Oxford, Keble Rd, Oxford OX1 3RH, UK
  \and
  Astronomy Program, Department of Physics and Astronomy,
Seoul National University, 1 Gwanak-ro, Gwanak-gu, Seoul 08826, Korea
  \and
  Department of Astronomy, Kyoto University, Kitashirakawa-Oiwake-cho, Sakyo-ku, Kyoto 606-8502, Japan
  \and
  Academia Sinica Institute of Astronomy and Astrophysics, 11F of Astronomy-Mathematics Building, AS/NTU, No.1, Section 4, Roosevelt Road, Taipei 10617, Taiwan
  \and
  Research Center for Space and Cosmic Evolution, Ehime University, 2-5 Bunkyo-cho, Matsuyama, Ehime 790-8577, Japan
  \and
  Korea Astronomy and Space Science Institute (KASI), 776 Daedeok-daero, Yuseong-gu, Daejeon 34055, Korea
  }

  \date{}

  \abstract
  {The North Ecliptic Pole (NEP) field provides a unique set of panchromatic data, well suited for active galactic nuclei (AGN) studies. Selection of AGN candidates is often based on mid-infrared (MIR) measurements. Such method, despite its effectiveness, strongly reduces a catalog volume due to the MIR detection condition.
  Modern machine learning techniques can solve this problem by finding similar selection criteria using only optical and near-infrared (NIR) data.}
  {Aims of this work were to create a reliable AGN candidates catalog from the NEP field using a combination of optical  SUBARU/HSC and NIR AKARI/IRC data and, consequently, to develop an efficient alternative for the MIR-based AKARI/IRC selection technique.}
  {A set of supervised machine learning algorithms was tested in order to perform an efficient AGN selection. Best of the models were formed into a majority voting scheme, which used the most popular classification result to produce the final AGN catalog. Additional analysis of catalog properties was performed in form of the spectral energy distribution (SED) fitting via the CIGALE software.}
    {The obtained catalog of 465 AGN candidates (out of 33 119 objects) is characterized by 73\% purity and 64\% completeness. This new classification shows consistency with the  MIR-based selection. Moreover, 76\% of the obtained catalog can be found only with the new method due to the lack of MIR detection for most of the new AGN candidates. Training data, codes and final catalog are available via the github repository. Final AGN candidates catalog is also available via the CDS service.}
  {New selection methods presented in this paper proves to be a better alternative for the MIR color AGN selection. Machine learning techniques not only show similar effectiveness, but also involve less demanding optical and NIR observations, substantially increasing the volume of available data samples.}
  

  \maketitle

\section{Introduction} \label{sec:intro}

Active galactic nuclei (AGN) are identified with centers of galaxies occupied by supermassive back holes (SMBH) in the phase of violent accretion of surrounding matter. Despite the known relation between the activity of central supermassive black holes and the overall evolution of galaxies, in particular star formation processes, our understanding of this co-evolution behaviour is still incomplete~\citep{netzer15unified_model_revised, padovani17}. 

An AGN definition can be assigned to a broad class of objects, from quasars, the most luminous objects in the universe, to spiral galaxies characterized by a high nuclear luminosity (Seyfert galaxies) or radio galaxies. These types display different observational properties in both optical and non-optical domains. This diverseness does not necessarily correspond to differences in the physical properties of objects and can be explained in terms of the AGN unified model \citep{antonucci93unified_model, netzer15unified_model_revised}. In the unified model, an AGN is described as an axisymmetric layered structure. It is formed by a central engine surrounded by the sub-pc accretion disc of infalling matter. At the range of $\simeq$ 1 pc, the broad line region (BLR) of high density, dust free gas is located. Outside a BLR lies a dusty torus, the geometry of which determines the ionization cone shape. The cone itself is a place of lower density ionized gas - a narrow line region (NLR). In its basic form, the unified AGN model, described by \citet{antonucci93unified_model}, attempts to characterize different AGN types as a combination of two parameters: the torus inclination to the line of sight and the source luminosity. 

Two main AGN types are referred to as type-I and type-II. Type-I AGN are characterized by the presence of broad permitted and semi-forbidden emission lines. Objects of this type with low to intermediate luminosity also show the presence of strong narrow emission lines. The appearance of a type-I is explained in terms of a geometrical phenomena - an AGN, with the ionization cone facing towards the observer, shows emission from the BLR as well as from the NLR (~\citeauthor{antonucci93unified_model}; \citealp{stern12typeI_I}). If the observer is not able to observe the BLR directly, then the AGN is defined as type-II. It is characterized by narrow emission lines showing photoionization features~\citep{merloni14agn_obscuration}. In this case the observer is facing a side of the AGN's dusty torus, so the emission from the BLR might be absorbed and reemitted at longer wavelengths in the infrared. Nonetheless, broad emission lines can still be observed via spectropolarimetrical techniques~\citep{marinucci12, tran03}. This two type classification is further complicated by the existence of so called "true type-II", with no detectable broad lines, and without a strong dust coverage~\citep{panessa02, shi10}. Moreover, some works (e.g. \citealp{assef15}) suggest an under-representation of type-II objects in a modern catalogs, causing a lack of realistic perspectives of the coexistence of both types.

Another big question on AGN physics is based on recently discovered changing look (CL) AGNs (e.g. \citealp{lamassa2015cl_agn, sheng2017cl_agn_mir, stern2018cl_agn_mir, charlton2019cl_agns}). This type of objects can exhibit transitions between AGN types by emergence or disappearance of broad emission lines. Generally accepted hypothesis about the mechanism of CL AGN activity, based on instabilities in the accretion disc, cannot be explained in terms of unified model and requires further research~\citep{lawrcence2018qso_viscosity_crisis, dodd2020cl_agn_landscape}.

\subsection{AGN photometric selection}

In order to better understand the physics hidden behind the AGN appearance one needs to gather a large amount of observational data in different spectral ranges. The AGN classification is commonly done via "diagnostic diagrams" describing emission and absorption line ratios (e.g. \citealp{baldwin1981PASP_bpt_diagrams, kauffmann2003MNRAS_classification_diagrams,kewley06agn_classification_via_diagrams}). Such classification requires time consuming spectroscopic measurements, which should be preceded by a careful target selection. 

A general characteristic of many AGN types is the form of a power-law shaped spectral energy distribution (SED), that allows one to effectively construct such preselection as a photometric classification task~\citep{alonsoherrero2006Apj_power_law_sed}. A power-law SED shape of an AGN, caused mainly by a central engine activity, shows the emission growth with the wavelength from the ultraviolet (UV) to infrared (IR) wavelength ranges~\citep{elivs1994ApJS_agn_sed_map, alonsoherrero2006Apj_power_law_sed}.
Despite this common property, AGN catalogs selected with photometry-only data obtained from various wavelength ranges can be biased and contaminated by other sources. 

Optical broad band selection is sensitive to broad emission lines. In the rest frame, optical photometry gives a standard set of colors providing easy way for AGN separation. However, due to varying redshift, the same AGN emission lines can be detected in different passbands, often giving stellar-like colors. This results in low completeness or high contamination in the redshift ranges between 2.6 and 3.5~\citep{richards2002AJ_sdss_spec_qso_target_selection}. 
Examples of an optical selection can be a quasar (QSO) target selection from the SDSS~\citep{york2000AJ_sdss} spectroscopic observations, where the combination of morphological properties and color selection methods have been used to separate QSOs from stars~\citep{richards2004ApJS_qso_selection_from_sdss_paperI, richards2009ApJS_sdss_qso_selection_paperII}. The biggest issue with the optical-only selection is a very low completeness of type-II AGN, caused by colors, which might be easily classified as stellar-like or type-I-like due to  conjunction of host and scattered AGN emission~\citep{zakamska2003AJ_typeII_in_vis, zakamska2019MNRAS_typeII_vis}.

A wavelength range that is free from the mentioned bias is the X-ray range. The hard X-ray emission, taking place in the accretion disk region (and also in the region of jets) can penetrate the dust and gas, giving an opportunity to detect even very obscured objects and complement the underrepresented type-II AGN class (e.g. \citealp{malizia2012MNRAS_x_vs_vis_obscuration}). The main selection effect present in X-ray AGN catalogs comes from a higher absorption of lower energy X-ray photons, which may interfere with the identification of heavily obscured AGN~\citep{padovani17}. Nevertheless, weak X-ray emission from a host galaxy as well as easily recognizable emission from a central engine makes X-ray surveys an efficient tool for AGN selection (e.g.~\citealp{lehmer2012ApJ_chandra4, luo2017ApJS_chadra7}). 

Another way to select AGN is to look for the infrared emission of a dusty torus, which comes from a reprocessed emission of accretion disk and, due to power-law behaviour, dominates AGN SED in the near- (NIR, 1-3~$\mu$m) and mid- (MIR, 3-50~$\mu$m) infrared. The NIR AGN selection (e.g.~\citealp{francis2004AJ_nir_agn_selection, kouzuma2010MNRAS_nir_agn_selection}), while being highly prone to contamination (see below), gives essential information about red AGN and red QSO in particular~\citep{glikman2007ApJ_nir_red_qso, glikman2012ApJ_nir_red_qso, glikman2013ApJ_nir_red_qso, banerji2012MNRAS_nir_red_qso}. More efficient and common is MIR AGN selection (e.g.~\citealp{deGrijp1987AAS_iras_agn, stern2005ApJmir_agn_selection, oyabu2011AA_agn_akari_mir_allsky, stern2012ApJwise_agn_selection}). It enables to fully take advantage of the power-law SED to detect both type-I and II AGN, that are characterized by hot dust MIR emission. Despite its effectiveness, MIR AGN selection is not free of biases and contamination sources which should be carefully studied (for further discussion see ~\citealp{padovani17}).

\textit{Contaminants}. One of the most informative parts of AGN emission is located in range of 3-5~$\mu$m, where the AGN stand out from the rest of the sources as typically redder objects. There are two main types of contaminants. Both share similar colors and should be taken into account during selection. Near the galactic plane, the main contamination comes from brown dwarfs (BD) and young stellar objects (YSO, \citealp{koenig2012ApJ_yso_in_mir}). Especially YSO, as a much more numerous class, can strongly reduce an AGN catalog purity. On high galactic latitudes, AGN can be easily confused with star forming galaxies (SFG), because AGN-like MIR colors can be produced both by polycyclic
aromatic hydrocarbons (PAH) molecules emission as well as the 1.6~$\mu$m stellar bump. At low-z, the stellar bump is a serious contamination source in NIR-selected AGN catalogs, at the same time high-z SFG occupy AGN color space in the MIR, forcing to set a completeness-purity trade-off~\citep{donley2010ApJ_sfg_contamination, assef2013ApJ_sfg_mir_contamination}.

\textit{Biases}. Bias present in MIR-selected AGN catalogs comes from three main sources: redshift dependence, dust luminosity and black hole scaling relations. Sensitivity of short MIR wavelengths to hot dust emission decreases with increasing redshift due to both K-correction and H$\alpha$ emission line contamination. The latter results in a strong bias against AGN at redshifts $z=[4, 5]$ (e.g.~\citealp{assef2013ApJ_sfg_mir_contamination}). Redshift-based biases present in IR and optical AGN selections can be partially bypassed by combined optical-IR methods~\citep{richards2015ApJS_qso_vis_and_mir, banerji2015MNRAS_nir_and_vis_agn_selection}. However, some AGN might also be missed in IR selection due to low dust content or due to low hot dust emission relative to emission from the host galaxy (see e.g.~\citealp{roseboom2013MNRAS_dust_emission_bias}). The last source of bias comes from the correlation of SMBH mass with luminosity of spheroidal component detected for low redshift AGN~\citep{marconi2003ApJ_bh_mass_scaling_relation}. This dependency relates the ratio the of AGN luminosity to host luminosity measured at a particular wavelength $(L_{\lambda , AGN}/L_{\lambda, host})$ to $L/L_{Edd}$, where $L_{Edd}$ refers to Eddington luminosity. Since the MIR AGN selection is sensitive only to the high value tail of the $L/L_{Edd}$ distribution, AGN with significant disk or irregular components need to show high $L/L_{Edd}$ value to be selected in the MIR~\citep{hickox2009ApJ_bh_mass_bias}. As a consequence, lower mass galaxies, which often have a strong non-bulge component, are underrepresented in MIR AGN catalogs~\citep{magorrian1998AJ_gal_mass_bh_mass}.

All selection methods discussed above primarily use color cuts and magnitude limits to define the parameter space occupied by AGN candidates and do not fully utilize advantages given by operating in a high dimensional space. Modern approaches to data analysis often use automated \textit{machine learning} (ML) techniques, which allow one not only to discover more sophisticated dependencies in a high dimensional space, but also to efficiently process big data volumes present in modern day surveys. An ML-based AGN selection is mostly done in a supervised manner, where an algorithm  first learns how to select a specific class of objects on the data set with known classes (we refer to this process as \textit{training}). After the learning stage, a trained algorithm can classify objects with unknown class (or label) during a \textit{generalization} stage. Supervised learning allows one to precisely control the process of selection at the expense of effectiveness which depends on a training sample quality. Supervised learning was successfully applied to AGN selection in broad a spectral range, e.g. in X-ray surveys~\citep{mcglynn2004ApJ_rosat_all_sky_multiclass}, optical data~\citep{claeskens2006MNRAS_qso_gaia_nn, nakoneczny2019AA_paperI} and IR or combined optical-IR data~\citep{zhang2004AA_clustering, malek2013AA_svm_vimos, nakoneczny2020paper2}.

\subsection{North Ecliptic Pole Field}
\label{sec:nep_intro}
A satisfactory picture of complex AGN properties requires information over a broad spectral range as described in the previous paragraphs. Hence, the most suitable for AGN studies are fields with broad panchromatic coverage. A region at the \textit{North Ecliptic Pole} (NEP; $\alpha=18^{h}00^{m}00^{s}$, $\delta=66^{\circ}33'88"$) is well suited for this task, since it provides broad range of information: from X-ray observations taken by ROSAT and Chandra space missions~\citep{henry06rosat, krumpe15chandra_nep_deep}, UV (GALEX satellite, \citealp{burgarella19galex_herschel}), optical measurements taken by Canada France Hawaii Telescope (CFHT) and, recently, by the SUBARU Hyper-Suprime Camera (HSC) instrument ~\citep{hwang07cfht, huang20cfht_megacam, ho20hsc_nep}, NIR observations performed by the CFHT/Wircam and FLAMINGOS instruments~\citep{oi2014A&A_cfht_wircam_nep_nir, jeon2014ApJS_flamingos_nep}, IR covered by the space missions AKARI, Spitzer Space Telescope, WISE and Herschel~\citep{kim12nepwide, jarrett11spitzer_wise, nayyeri18spitzer, pearson2017PKAS_herschel}, up to sub-mm (SCUBA-2, \citealp{geach17scuba2, shim2020MNRAS_scuba2}) and radio (WSRT, GMRT) wavelengths~\citep{ white10wsrt20cm, white17gmrt614MHz}. The high galactic latitude of the NEP field ensures little galactic dust and star pollution, creating a good conditions for deep, extragalactic observations. 

The NEP panchromatic data have been extensively used to test and/or develop several AGN selection techniques, such as SED fitting classification~\citep{huang2017MNRAS_18band_sed_agn_selection_nep_deep, oi2020MNRAS_hsc, Wang2020, yang2020MNRAS_xcigale}, traditional color selection~\citep{jarrett11spitzer_wise, fadda2014MNRAS_mir_color_selection_nep_and_others}, combinations of both~\citep{barrufet20AA_nep_high_z_galaxies}, machine learning selection~\citep{poliszczuk19, chen2021MNRAS_nn} or radio-based selection~\citep{karouzos2012PKAS_radio_agn_selection_nep, barrufet2017PKAS_vis_radio_nep_agn}.

\subsection{Outline}

This work aims to create a new, reliable AGN catalog from NEP data based on a combination of optical data from SUBARU/HSC and NIR data gathered by AKARI satellite. The AKARI satellite mission~\citep{murakami07akari} carried out observations in near-(NIR) mid-(MIR) and far-infrared (FIR). In particular it observed the NEP field with NIR and MIR passbands in two surveys~\citep{matsuhara2006PASJ_akari_nep_surveys}: AKARI NEP-Wide~\citep{lee2009PASJ_nep_wide, kim12nepwide}, and AKARI NEP-Deep~\citep{wada2008PASJ_akari_nep_deep}, creating unique IR catalogs in terms of both, photometric coverage and depth. Data gathered by the AKARI telescope in conjunction with new observations of the NEP field provided by the SUBARU/HSC instrument~\citep{oi2020MNRAS_hsc, seongjin20_new_catalog} establish the cutting edge environment for AGN selection as well as setting the broad picture of its optical-IR relation. This data set can be also viewed as a test ground for upcoming synergies of the future photometric surveys, such as Vera C. Rubin Observatory~\citep{ivezic19lsst} and Euclid~\citep{laureijs10euclid}. 

A ML approach to combined NIR and MIR AGN selection in AKARI NEP-Deep data described in~\citet{poliszczuk19} serves as an introductory study for the present paper in terms of feature importance, extrapolation risk estimation and application of fuzzy logic to classification algorithms. The new method overcomes previously encountered  difficulties related to small generalization volume and extrapolation risk and 
shows that it is possible to recover properties of AGN MIR selection using only optical and NIR data. Such improvement significantly increases the catalog volume and makes the presented method suitable for modern AGN studies. The final AGN catalog, training sample, Python codes and training results can be found on authors github~\footnote{\url{https://github.com/ArtemPoliszczuk/NEPWide_AGN}}. The AGN catalog can be also accessed via CDS database and Virtual Observatory tools~\footnote{\url{https://ivoa.net/}}.

This present work is organized as follows. Sec.~\ref{sec:data} describes the data used for training and generalization stages of ML. Sec.~\ref{sec:method} walks through the applied ML pipeline. Sec.~\ref{sec:results} describes the classification evaluation results, obtained AGN catalog, and comparison with other AGN selection methods. Sec.~\ref{sec:summary} summarizes properties of the obtained catalog as well as effectiveness of the applied methods. Appendix contains information about software used in the present work and additional data describing the performance evaluation.

\section{Data} \label{sec:data}

\subsection{AKARI NEP-Wide data}

The AKARI NEP-Wide survey covers a circular area of 5.4~deg$^2$ centered at the North Ecliptic Pole (NEP). It provides photometric observations in the NIR and MIR passbands, taken by the Infra-red Camera instrument (IRC, \citealp{onaka07irc}). The IRC was equipped with nine filters: three NIR filters centered at 2~$\mu$m (N2; with coverage range 1.9-2.8~$\mu$m), 3~$\mu$m (N3; 2.7-3.8~$\mu$m), 4~$\mu$m (N4; 3.6-5.3~$\mu$m) and six MIR filters centered at 7~$\mu$m (S7; 3.6-5.3~$\mu$m), 9 $\mu$m (S9W; 6.7-11.6~$\mu$m), 11~$\mu$m (S11; 8.5-13.1~$\mu$m), 15~$\mu$m
(L15; 12.6-18.4~$\mu$m), 18~$\mu$m (L18W; 13.9-25.6~$\mu$), and 24~$\mu$m (L24; 20.3-26.5~$\mu$m). The "W" letter next to filters centered at 9~$\mu$m and 18~$\mu$m refers to wide coverage of wavelengths relevant to the neighbouring filters (e.g. S9W covers the wavelength range of S7 and S11 filters).

The IRC continuous wavelength coverage as well as choice of filter placement and profile gives important insights into the stellar and AGN activity (for further discussion see ~\citealp{matsuhara2006PASJ_akari_nep_surveys}). The N2 and N3 filters can trace the stellar mass of galaxies at different redshifts as well as help to identify star-forming galaxies (SFG) due to the location of the 1.6~$\mu$m stellar bump. Moreover, the NIR filters are sensitive to the 3.3~$\mu$m PAH feature of local starburst galaxies and carbon dust absorption feature located at 3.4~$\mu$m, which characterizes low-z obscured AGN. The N4 passband is the most crucial from the perspective of AGN identification, since it covers 3-5~$\mu$m range, where the power-law properties of an AGN SED are the most prominent. The 9-20~$\mu$m range covered by MIR passbands gives the possibility to trace dust emission from AGN and LIRG/ULIRG due to presence of several silicate dust features with particularly strong 9.8 $\mu$m absorption features. On the other hand, PAH emission features centered around 7.7~$\mu$m fall into the range of the S7 band, allowing identification of starburst galaxies~\citep{kim2019PASJ_MIR_PAH_emission_SFG}.

\subsection{Optical photometry and training sample}
\label{sec:data_training}

SUBARU/HSC~\citep{miyazaki12hsc} optical observations of the NEP field~\citep{goto2017PKAS_hsc_nep} allows one to detect optical counterparts for even the faintest of AKARI sources. Previous optical follow-ups of the AKARI NEP field performed by CFHT/MegaCam~\citep{hwang07cfht} and Maidanak/SNUCam~\citep{jeon_2010ApJS_maidanak_nep} instruments were unable to detect sufficient fraction of IR sources due to the depth limit. New HSC data processed by~\citet{oi2020MNRAS_hsc} and matched with AKARI data by~\citet{seongjin20_new_catalog} gives measurements in $g$, $r$, $i$, $z$, $Y$ passbands $1.7\mkern2mu{-}\mkern2mu 2.5$~mag deeper than previous surveys.

Optical and IR photometry gathered by SUBARU/HSC and AKARI/IRC was used to perform AGN selection on the data set. Additionally, one needs class labels for a training sample in order to adjust ML algorithms to a specific classification task. The majority of labels for a training sample were taken from spectroscopic follow-up~\citep{shim13nepspec} performed by MMT/HECTOSPEC~\citep{fabricant05hectospec} and WYIN/HYDRA~\citep{barden93hydra} spectrographs. The choice of training sample is crucial in ML classification methods since it has direct impact on the algorithm performance and generalization possibilities. Hence, both target selection criteria as well as the process of class assignment in the labeled data must be well understood. Primary targets for spectroscopic observations were defined as objects bright in MIR passbands ($S11\mkern2mu{<}\mkern2mu18.5$~mag and $L15\mkern2mu{<}\mkern2mu17.9$~mag) with an additional limit in the Maidanak $R$ band ($16\mkern2mu{<}\mkern2mu R\mkern2mu{<}\mkern2mu 21{-}22.5$~mag depending on a spectrograph) in order to select objects suitable for spectroscopic observations. A secondary targets group was made of specific classes of candidates described in terms of optical, NIR and MIR color cuts. AGN candidates from the secondary target group were selected using a color cut method developed by~\citet{lee07}, which uses power-law SED properties of AGN manifested in red NIR and MIR colors: 
$N2\mkern2mu{-}\mkern2mu N4\mkern2mu{<}\mkern2mu 0$ 
and $S7\mkern2mu{-}\mkern2muS11\mkern2mu{<}\mkern2mu0$. This method was originally used on the $S11\mkern2mu{<}\mkern2mu18.5$~mag limited sample. The catalog, presented in~\citet{shim13nepspec}, was augmented by additional spectroscopic measurements taken by Keck/DEIMOS~\citep{faber03deimos}, GTC/OSIRIS~\citep{cepa00osiris} and SUBARU/FMOS~\citep{kimura10fmos} instruments by AKARI NEP Team as well as X-ray measurements from the Chandra~\citep{weisskoph00chandra, krumpe15chandra_nep_deep} telescope. This way the AGN training sample was constructed mainly from targets fulfilling the \citet{lee07} AGN candidate conditions or X-ray active sources. Sources detected by Chandra with an X-ray luminosity of $\log{L_X}\mkern2mu{>}\mkern2mu 41.5$~erg/s (0.5-7~keV) will be referred to as XAGN. Optically detected objects with at least one emission line with Full Width at Half Maximum (FWHM) larger than 1000 km/s are referred to as AGN1.

Tab.~\ref{tab:num_of_objects} shows the number of objects measured in particular HSC and IRC passbands. 
The decrease in the data volume for the passbands corresponding to the longer wavelengths is caused by  decreasing sensitivity in the MIR range as well as the gradual disappearance of Galactic stars in the MIR bands. 
In order to preserve larger data volume only optical and NIR data were used for the AGN selection in the present work. It resulted in 1 547 objects in the training set consisting of 1 348 galaxies and 199 AGN (163 AGN1 and 36 XAGN) and 45 841 in the generalization set (additional limitations applied to the generalization set are described in Sec.~\ref{sec:gen_lim_method}). Tab.~\ref{tab:training_stat} shows general statistical properties of the training sample.

Fig.~\ref{fig:n4_n2n4_train} presents a $N2\mkern2mu{-}\mkern2muN4$ vs $N4$ color-magnitude (CM) plot for training data. This particular CM dependency is often used for AKARI/IRC data presentation \cite[e.g. ][]{lee07, lee2009PASJ_nep_wide} and is present in classification analysis in Sec.~\ref{sec:results}. The $N2\mkern2mu{-}\mkern2mu N4$ color, being sensitive to power-law propreties of the short MIR AGN SED, brings out the main characteristics of the selection task described in the present paper. One can see that the main locus of type-I AGN is placed in the red part of $N2\mkern2mu {-}\mkern2mu N4$ color as it is imposed by the selection creteria of the~\cite{lee07} method. One can also observe the increase of galaxy redshift with $N2\mkern2mu{-}\mkern2mu N4$ reddening, making high-z SFG the main contaminant source in the type-I AGN area. XAGN objects are scattered around whole training sample area, reaching even very low $N2 \mkern2mu {-}\mkern2mu N4$ values. The stellar locus, not represented in this work is placed in the left lower corner of the plot.

\begin{table}[t] 
\centering
\begin{tabular}{lll} 
\textbf{band} & \textbf{catalog size}  & \textbf{labeled data} \\
\hline 

g &         \textbf{89 835}         &         1 870  \\
r &         \textbf{89 431} ( 88 642) &         \textbf{1 869} ( 1 867) \\
i &         \textbf{87 385} ( 86 186) &         \textbf{1 864} ( 1 860) \\
z &         \textbf{89 028} ( 86 023) &         \textbf{1 871} ( 1 859) \\
Y &         \textbf{86 587} ( 84 874) &         \textbf{1 862} ( 1 856) \\
N2 &        \textbf{61 679} ( 59 845) &         \textbf{1 650} ( 1 637) \\
N3 &        \textbf{74 475} ( 54 152) &         \textbf{1 743} ( 1 598) \\
N4 &        \textbf{66 134} ( 45 841) &         \textbf{1 722} ( 1 547) \\
S7 &          \textbf{5 041} ( 4 168) &         \textbf{998} (918) \\
S9 &          \textbf{9 316} ( 3 536) &         \textbf{1 360} (882) \\
S11 &         \textbf{9 147} ( 3 167) &         \textbf{1 320} (843) \\
L15 &         \textbf{8 688} ( 2 404) &         \textbf{1 070} (729) \\
L18 &         \textbf{10 258} (2 294) &         \textbf{1 131} (704) \\
L24 &         \textbf{2 450} (1 208) &          \textbf{520} (437) \\

\hline
\end{tabular}
\caption{Number of objects detected in particular SUBARU/HSC and AKARI/IRC passbands. Numbers in brackets show sources with  measurements existing in all previous passbands corresponding to shorter wavelengths.}
\label{tab:num_of_objects}
\end{table}

\begin{table}[t]
    \centering
    \begin{tabular}{lllll}
    
    {} &  \textbf{median} & \textbf{MAD} & \textbf{min.} & \textbf{max.} \\
    \hline 
    redshift &   0.339 &  0.308 &  0.001 &   4.320 \\
    g     &  21.075 &  1.313 &  16.224 &  27.109 \\
    r     &  20.126 &  1.188 &  15.594 &  26.264 \\
    i     &  19.610 &  1.101 &  15.254 &  25.214 \\
    z     &  19.296 &  1.066 &  15.056 &  24.781 \\
    Y    &  19.119 &  1.059 &  14.850 &  24.278 \\
    N2    &  18.543 &  0.859 &  14.079 &  20.814 \\
    N3    &  18.692 &  0.732 &  14.528 &  20.638 \\
    N4    &  18.951 &  0.711 &  15.007 &  20.935 \\

    \hline
    \end{tabular}
    \caption{Statistical properties of the training sample. Median, median absolute deviation (MAD), minimal and maximal values for redshift and passbands used for training are shown.}
    \label{tab:training_stat}
\end{table}

\subsection{Panchromatic NEP catalog}

In the present work properties of both training data set and the obtained catalog were investigated a posteriori via SED fitting analysis. In order to obtain a robust estimation of the SED, one needs a broad wavelength photometric coverage of the studied object. For this purpose, part of the~\citet{seongjin20_new_catalog} panchromatic catalog covering optical to sub-mm wavelength range was used. In order to obtain more precise fitting results and to ensure the correct matching of objects from different surveys, several limitations were made in the further analysis: the WISE $W4$ data were excluded, $g$, $r$, $i$, $z$ and $Y$ measurements were taken only from SUBARU/HSC instrument and the $J$ passband data were taken from CFHT/Wircam instead of FLAMINGOS, despite the smaller field observed by the first instrument.

\section{Methodology} \label{sec:method}

\pgfdeclarelayer{background}
\pgfdeclarelayer{foreground}
\pgfsetlayers{background,main,foreground}

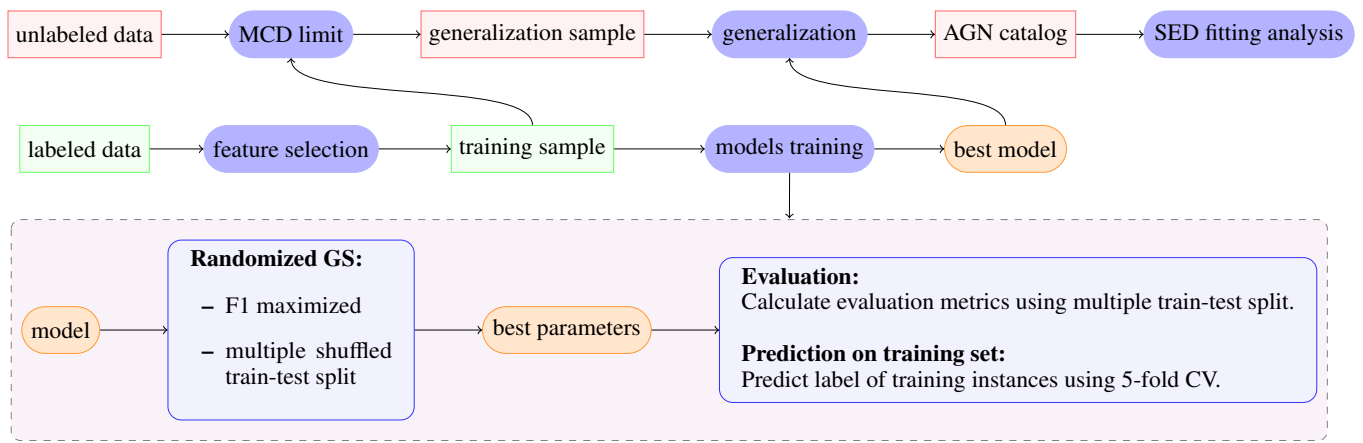
\begin{figure*}
\centering
\begin{adjustbox}{width=7in}
\begin{tikzpicture}[
traindatanode/.style={rectangle, draw=green!60, fill=green!5, thin, minimum size=7mm},
generaldatanode/.style={rectangle, draw=red!60, fill=red!5, thin, minimum size=7mm},
operationnode/.style={rounded rectangle, draw=blue!30, fill=blue!30, thin, minimum size=7mm},
modelnode/.style={rounded rectangle, draw=orange!80, fill=orange!20, thin, minimum size=7mm},
trainoperationnode/.style={rectangle, draw=blue!80, fill=blue!5, thin, minimum size=7mm, rounded corners=2mm}
]
\node[generaldatanode] (unlab) {unlabeled data};
\node[operationnode] (mcd) [right=of unlab] {MCD limit};
\node[generaldatanode] (general) [right=of mcd] {generalization sample};

\node[traindatanode] (lab) [below=of unlab] {labeled data};
\node[operationnode] (featureselect) [below=of mcd] {feature selection};
\node[traindatanode] (traindata) [below=of general] {training sample};

\node[operationnode] (generalization) [right=of general] {generalization};
\node[operationnode] (training) [below=of generalization] {models training};

\node[generaldatanode] (agn) [right=of generalization] {AGN catalog};
\node[operationnode] (sed) [right=of agn] {SED fitting analysis};

\node[modelnode] (clf) [below=of agn] {best model};

\draw[->] (unlab.east) -- (mcd.west);
\draw[->] (mcd.east) -- (general.west);

\draw[->] (lab.east) -- (featureselect.west);
\draw[->] (featureselect.east) -- (traindata.west);

\draw[->] (traindata.north) .. controls +(up:7mm) and +(down:7mm) .. (mcd.south);

\draw[->] (traindata.east) -- (training.west);
\draw[->] (training.east) -- (clf.west);
\draw[->] (general.east) -- (generalization.west);
\draw[->] (clf.north) .. controls +(up:7mm) and +(down:7mm) .. (generalization.south);

\draw[->] (generalization.east) -- (agn.west);
\draw[->] (agn.east) -- (sed.west);


\node[trainoperationnode] (gs) [below=of featureselect]{
\begin{tabular}{l}
  \textbf{Randomized GS:}\\
  \parbox{3cm}
  {
  \begin{itemize}
   \item F1 maximized
   \item multiple shuffled train-test split
  \end{itemize}
  }
\end{tabular}
};

\node[modelnode] (model) [left=of gs] {model};

\node[modelnode] (params) [right=of gs] {best parameters};

\node[trainoperationnode] (eval) [right=of params]{
\begin{tabular}{l}
  \textbf{Evaluation:}\\
  Calculate evaluation metrics using multiple train-test split. \\ \\
  \textbf{Prediction on training set:} \\
  Predict label of training instances using 5-fold CV.
\end{tabular}
};

\draw[->] (model.east) -- (gs.west);
\draw[->] (gs.east) -- (params.west);
\draw[->] (params.east) -- (eval.west);

\begin{pgfonlayer}{background}
        \path (model.west |- gs.north)+(-0.15,0.3) node (a) {};
        \path (eval.east |- gs.south)+(+0.15,-0.3) node (b) {};
        \path[fill=violet!5,rounded corners, draw=black!50, dashed]
            (a) rectangle (b);
\end{pgfonlayer}

\draw[->] (training.south) -- (10.465, -2.75);

\end{tikzpicture}
\end{adjustbox}
\caption{Scheme of the machine learning pipeline described in the present work. Upper part of the scheme shows the general outline, lower part of the scheme, shown in the violet rectangle, refers to the models training.}
\label{fig:pipeline}
\end{figure*}

\subsection{Feature selection} \label{sec:feature_selection_method}

Usual astronomical parameters are frequently not optimal for a specific ML task. To obtain a more suitable data representation one can use \textit{feature engineering} techniques for data transformation. In the present paper, the \textit{Kolomogorov-Smirnov two sample test}~\citep{massey51} was used. By calculating the biggest difference in cumulative distribution functions of two samples (referred to as the \textit{KS-statistic}) one can test if samples come from the same distribution. In a case of binary classification problem, this method can be used for feature selection by calculating the KS-statistic value for class samples represented by different features. Features, which will obtain the highest KS-statistic value, will show 
the biggest difference in class distributions, hence will be the most suitable for a specific classification problem. The KS two-sample test gives information about separate features, without estimating their importance in interaction with a particular feature subset. 

Another important issue that should be addressed during feature selection is the number of selected features. This number cannot be too high due to two (often connected) possible risks: first, one might introduce redundant features, that carry information already present in a feature set, second, the \textit{curse of dimensionality}~\citep{bishop06book} might cause substantial sparsity of the data in a high dimensional space resulting in a poor performance. In the present work, the limit imposed on the number of selected features was based on the assumption that adding a filter to the feature set should not introduce redundant information. Thus the optimal feature set should contain information from most of the available filters but the number of features should not be significantly larger than the number of available filters.

\subsection{Classification algorithms} \label{sec:algorithms_method}

In the present paper, several types of supervised classification algorithms were tested. The imbalanced nature of the training data set can have a major impact on the classifier performance and thus has to be taken into account at several stages of the ML pipeline construction. It is considered a good practice, to test several types of classification algorithms in a classical setup and in a setup suitable for imbalanced data~\citep{fernandez18}. It is done not only to find the best algorithm for a particular data set and classification task, but also to control how the growing model complexity will improve performance on the labeled data set. This issue can be analyzed in terms of bias-variance dependency of the model. A \textit{bias} of the model is the strength of assumptions it is making about a decision boundary. In general, linear models have a larger bias than nonlinear ones. If a bias is too big, a model will not be able to properly adapt to the training data. Such a model is referred to as \textit{under-fitted}. On the other hand, a model can learn specific training data too precisely, including the statistical noise of the sample. This situation is called \textit{overfitting}. An overfitted model will have high \textit{variance}, i.e. it will be very sensitive to changes in a training data. In an ideal situation, one should strive for an optimal bias-variance trade-off, where a model can be effectively fitted to a training data set without learning its random fluctuations.

Before applying real world algorithms, one has to settle on some lowest boundaries for a classifier performance comparison. For this purpose, a so called \textit{dummy classifier} was constructed. It classifies objects according to the class fraction in the training data. Next, a linear model with the form of the \textit{logistic regression} algorithm~\citep{berkson44} was tested. It relates a linear combination of the input variables (features) to the logarithm of the odds for the positive value target variable. This way it can directly model the probability of a particular target value for a given input. 

Often the separation of the classes in the input feature space is very hard or even impossible. The main idea of the \textit{support vector machine} (SVM) algorithm developed by~\citet{vapnik95} is to map data into the high dimensional space, where one can construct the separating hyperplane. An output of the SVM classifier relies on a position of objects with respect to such a hyperplane. This distance also serves as a basis for indirect probability estimation. In the SVM formulation the mapping from an input to a high-dimensional space is substituted by a kernel function which is set to be a radial basis function (RBF) in this work. 

Another family of popular classification algorithms are based on decision tree structures ~\citep{breiman84}. Single decision classifiers can perform very well on the training data, but due to the high variance of the model it often tends to overfit the data. An effective way to reduce the variance of the model is to introduce randomization into the training or construction of the classifier and then collect such different models into an ensemble. A final prediction is made by averaging the outputs of a particular classifiers. The \textit{random forest}~\citep{breiman01} algorithm incorporates two forms of randomness into the training. First, it trains each decision tree model on a different data set drawn with replacement from the training data. Secondly, the best split in each node is found using a random subset of features. In this case the number of features is controlled by a parameter that can be tuned. An additional level of randomness is used in the \textit{extremely randomized tree}~\citep{geurts06}, where node splitting is done not by the most discriminative threshold, but by setting a threshold at random for each candidate feature separately and taking the best threshold as the splitting rule. 

Besides averaging the output of particular classifiers, as it was described in the previous paragraph, one can build classifiers in a sequential form, where the model is trying to correct the output of its predecessor. This way of constructing ensembles is called \textit{boosting}. In recent years, the most popular and effective implementation of boosting tree ensembles is an \textit{XGBoost}~\citep{chen16} algorithm, which shows leading results in many ML competitions. 

In the case of an imbalanced data, a classifier often tends to minimize the area of feature space assigned to the smaller class prediction. As a consequence, one would obtain a catalog of smaller class candidates characterized by a high purity and low completeness. In order to minimize this tendency, one can introduce the \textit{class balance}, where the importance of smaller class objects is increased in training proportionally to the class fraction in the labeled data. All previously mentioned classifiers (except the dummy classifier) we tested in both class-balanced and non-balanced versions.

In most cases, the default values of the model hyperparameters are not optimal for a specific ML task. This issue was addressed in the present work by tuning hyperparameters via \textit{randomized grid search}. It samples a prearranged hyperparameter grid a certain number of times, giving a set of hyperparameter combinations used in a training. The effectiveness comparison of these combinations was done via performance evaluation described in Section~\ref{sec:evaluation_method}. While logistic regression, SVM and XGBoost classifiers were tuned using the above  method, random forest and extremely randomized trees algorithms were excluded due to their insensitivity to the tuning process~\citep{probst19rftuning}. 

Finally, having predictions from different models described above, one can combine their outputs into a voting schemes. In the present work two types of voters were used: \textit{hard voter classifier}, where the majority predicted class was used as a final result and the \textit{stacked classifier}, where the logistic regression was trained on the probability estimations from different models as features.

\subsection{Fuzzy logic} \label{sec:fuzzy_method}

\begin{figure*}[t]
\begin{subfigure}{.5\textwidth}
\centering
\includegraphics[width=0.8\textwidth]{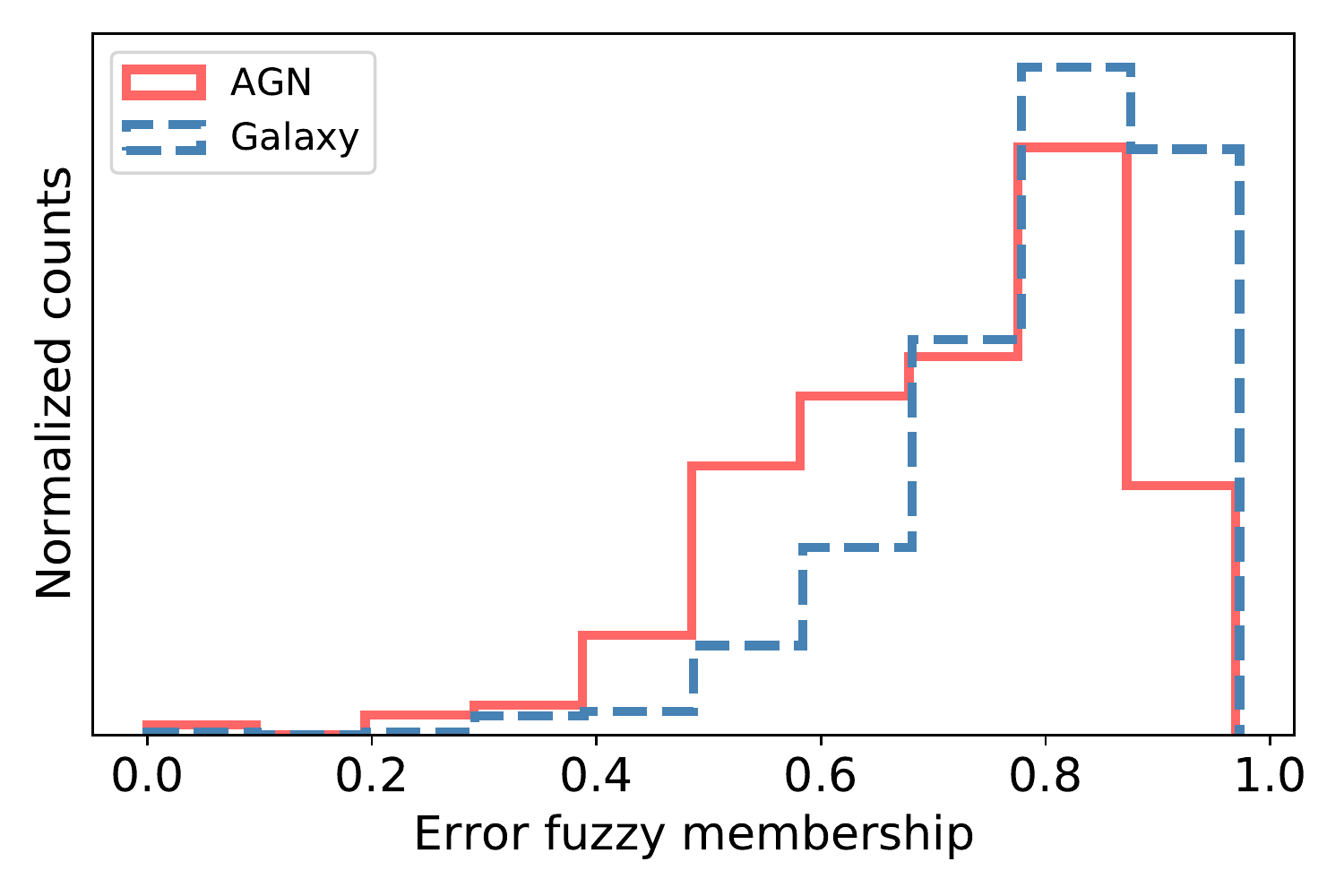}
\caption{Error-based fuzzy membership}
\label{fig:fuzzy_err}
\end{subfigure}
\begin{subfigure}{.5\textwidth}
\centering
\includegraphics[width=0.8\textwidth]{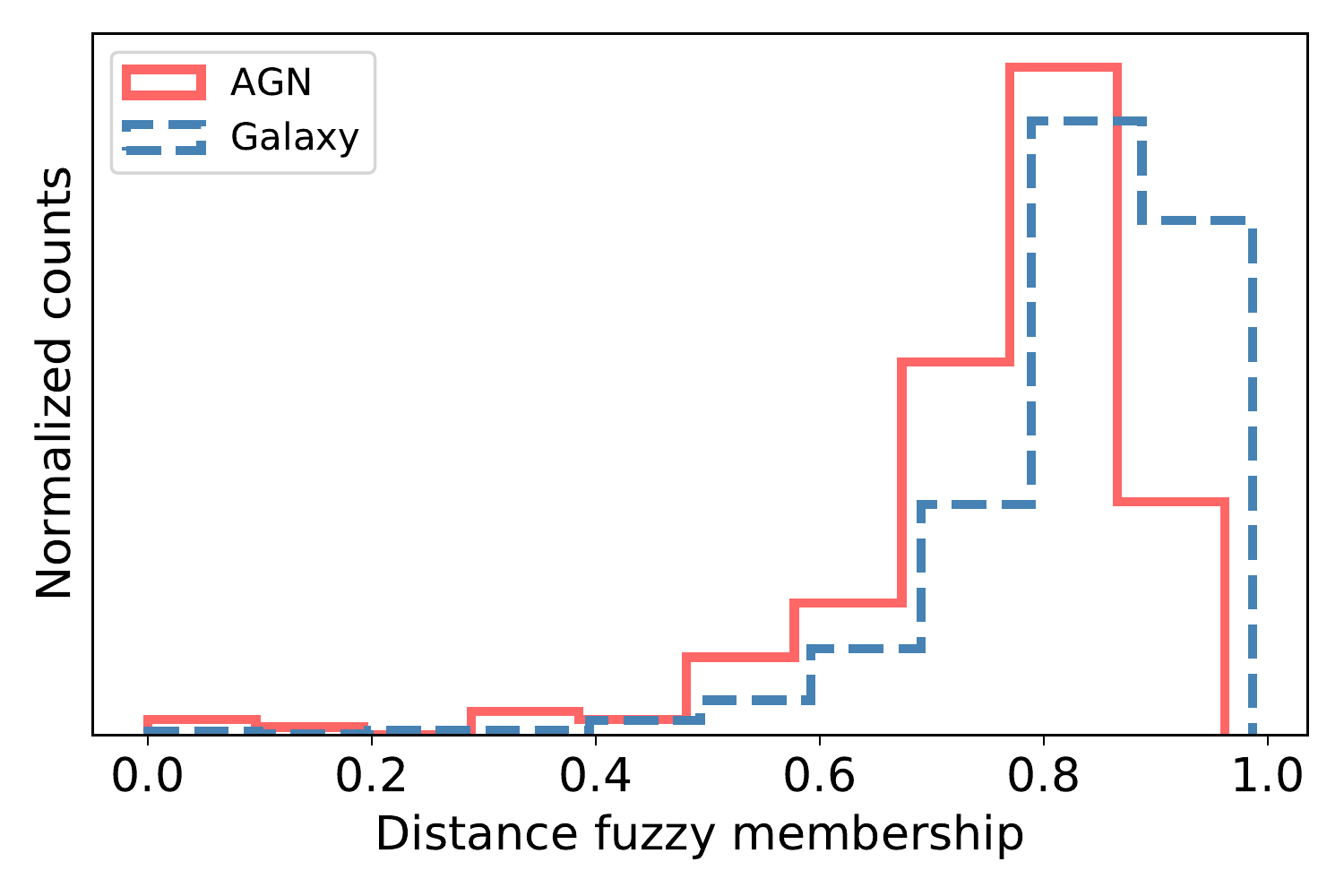}
\caption{Distance-based fuzzy membership}
\label{fig:fuzzy_dist}
\end{subfigure}
\caption{Fuzzy memberships calculated for the training data from first iteration (see Sec.~\ref{sec:first_iteration_results}). Error-based  weights were calculated with respect to optical and NIR passbands measurement uncertainties, giving higher priority to objects with better measurements within AGN or galaxy sample. Distance-based weights were calculated with respect to the class center calculated in a feature space, reducing impact of outliers on classification result.}
\label{fig:fuzzy_memberships}
\end{figure*}

One can further tune the weights of the labeled data to improve the training process by applying instance-weighting often referred to as \textit{fuzzy logic}. In this case each object in a training set has its own importance according to some chosen properties. This instance weight is referred to as \textit{fuzzy membership}. Specific applications of instance-weighting to logistic regression, SVM and tree-based algorithms are discussed in~\citep{hosmer13fuzzy, lin02fuzzy, ting02fuzzy} respectively.

In previous work on the SVM-based AGN selection~\citep{poliszczuk19}, authors used the measurement uncertainty based weighting system (referred to as \textit{error fuzzy} weights), where precisely measured objects were treated as more important than the noisy ones during training. It allowed authors to reduce the impact of the noise and increase the reliability of the output catalog.  In the present paper, besides the error weights, distance-based weights inspired by~\citet{lin02fuzzy} were also used. In this case the Euclidean distance in the feature space was translated to the importance weight of the object. This way objects typical for a particular class were treated as more important, while the influence of the outliers was reduced. In both cases fuzzy memberships $s_i$, normalized to the range [0,~1], were calculated using the same scheme:

\begin{equation}
    s_i = 1 - \dfrac{u_i}{u_{\text{max}} + \delta},
\label{eq:fuzzy}
\end{equation}

\noindent
where the $u_i$ is a measurement uncertainty of the object or its Euclidean distance from its class center and $u_{\text{max}}$ is a maximal uncertainty or distance for a specific class. The $\delta$ parameter is a small value used to avoid division by zero. 
Since only the relative difference between instance weights has an impact on the training process, the delta parameter is added only for a purpose of the numerical safety. In this work $\delta=10^{-4}$. Histograms of both, error- and distance-based fuzzy memberships calculated for galaxy and AGN training samples used for the creation of final catalog (first iteration, see Sec.~\ref{sec:results}) are shown in Fig.~\ref{fig:fuzzy_memberships}. Error-based weights were calculated with respect to measurement uncertainties of optical and NIR passbands. Distance-based weights were calculated with respect to class center in the high-dimensional space of selected features.

\subsection{Evaluation} \label{sec:evaluation_method}

The quality of classifier performance and obtained AGN catalog can be measured in two ways. The first approach is based on a comparison with results given by other methods - this type of evaluation is described in the next sections. The second approach is based on estimating the generalization performance by evaluation metrics calculation for the classification performed on the labeled data. In order to get a reliable evaluation one has to test the classifier on data that was not used in the training, since training scores, being prone to overfitting, might show inflated results. To address this issue, training-test splitting of the labeled data was done: the classifier was trained on a training part of the data and evaluated on a test one. This process was repeated many times by shuffling data before each training-test split. Such an approach allows one to get a meaningful evaluation with a measurement uncertainty.

Imbalanced data learning introduces additional challenges in the performance evaluation process. Simple metrics such as accuracy cannot be treated as reliable performance measures anymore due to their tendency to return high values even for poor performances on a smaller class~\citep{fernandez18}. The purpose of the present work is to produce a reliable catalog of AGN characterized by high purity and completeness, hence these should be the main metrics of a classifier performance evaluation. If we refer to the AGN class as \textit{positive} and galaxies class as \textit{negative}, the purity of the AGN catalog will be known as \textit{precision}:

\begin{equation}\label{eq:precision}
    \text{Precision} = \dfrac{T_p}{T_p + F_p},  
\end{equation}

\noindent
which is a fraction of properly classified AGNs (true positives $T_p$) to all AGN candidates (true positives and false positives $F_p$). On the other hand, completeness of the AGN catalog, referred to as \textit{recall} is a fraction of properly selected AGNs to all AGNs (true positives and false negatives) in the classified data and can be written as:

\begin{equation}\label{eq:recall}
    \text{Recall} = \dfrac{T_p}{T_p + F_n}. 
\end{equation}

\noindent
In order to maximize both these measures during learning and to effectively compare performance of different classifiers one can use several compound metrics. First of them is the F1 score which is a harmonic mean of precision and recall:

\begin{equation}\label{eq:f1}
    \text{F1} = 2 \dfrac{\text{Precision} \cdot \text{Recall}}{\text{Precision} + \text{Recall}}.
\end{equation}

\noindent
If one can obtain a probability estimation for a particular object belonging to one of two classes, the \textit{precision-recall curve} can be constructed. It is done by calculating the precision and recall scores for different decision thresholds. Beside analyzing the shape of the PR-curve one can evaluate the effectiveness of the classifier by calculating area under the PR-curve (\textit{PR-AUC}). 

The last metric used in this paper is the \textit{balanced accuracy} (bACC), which normalizes true positive and true negative predictions by the number of positive and negative samples, respectively. If we create a metric called the \textit{true negative rate} (TNR) which is the same as precision, but made for a negative class, we can express bACC as:

\begin{equation}\label{eq:bACC}
 \text{bACC} = \dfrac{1}{2}(\text{Precision}+\text{TNR}).
\end{equation}

\subsection{Generalization sample limit} \label{sec:gen_lim_method}

The supervised classification result can only be as good as the provided labeled sample is. This statement can be also expressed differently: a classifier can give reliable results only in areas of feature space occupied by training data. In other words one has to be very careful with extrapolation during the generalization phase, since we cannot control performance in the unknown regions of the feature space. Application of simple magnitude cuts, to limit the generalization data to the labeled data ranges is not enough, since a high-dimensional manifold created by generalization set in feature space might still preserve some regions requiring extrapolation.

In the present work a simple conservative approach of a \textit{minimum covariance determinant} estimator algorithm (MCD,~\citealp{rousseeuw99}) was used. It allows one to fit an n-dimensional ellipsoid to the training data and limit generalization data to its range. The MCD algorithm has a free parameter $\alpha$ called the \textit{contamination rate}, which defines the number of outliers present in a training sample and controls how conservative the limiting process will be.

In order to find adequate $\alpha$ value, the \textit{Mahalanobis distance} histograms were used. The Mahalanobis distance $d_M$ is defined as:

\begin{equation}
    d_M(\vec{x}) = \sqrt{(\vec{x}-\vec{\mu})^{T}\Sigma^{-1}(\vec{x}-\vec{\mu})},
\end{equation}

\noindent
where $\vec{x}$ is an object location, $\vec{\mu}$ is the mean and $\Sigma$ is the covariance matrix. Making a $d_M$ histogram one can look for the smallest $d_M$ value, corresponding to the histogram discontinuity. This range might be used to select an optimal $\alpha$ value.

\subsection{SED fitting}
\label{sec:SED_method}

Considering the tight link between the galaxy SED shape and the presence of an AGN, SED fitting was performed for the training sample and objects selected during generalization step.
The broad band multiwavelength SED fitting tool, The Code Investigating GALaxy Emission\footnote{\url{https://cigale.lam.fr/}}  \cite[CIGALE,][]{Boquien2019} was used to estimate the possible presence of an AGN and main physical parameters of the sample.

The CIGALE software was incorporated into the analysis, as it is a  physically-motivated state-of-the-art python code for SED fitting of the extragalactic sources, and  moreover, it was already used and tested for the AKARI-NEP galaxies \cite[i.e.][]{Buat2015,Solarz2015,Toba2020,barrufet20AA_nep_high_z_galaxies} and AGN (see Sec.~\ref{sec:nep_intro}). This code employs both galaxy and AGN models and allows one to fit both components in a flexible way.
CIGALE simultaneously fits the AGN and galaxy spectrum from far-UV to far-IR\footnote{The special version of CIGALE, called X-CIGALE, allows to fit SEDs from the X-ray to far-IR, \cite{yang2020MNRAS_xcigale}.}  and returns the estimate of the main galaxy properties such as stellar mass, dust luminosity, star formation rate, and the relative contribution of the dusty torus of the AGN to the total IR luminosity - the AGN fraction. 
For each estimated parameter CIGALE uses a probability distribution function (PDF) analysis, and the final physical parameters are built based on the likelihood-weighted mean of the PDF. 
Each parameter is associated with an error calculated also from the  PDF's likelihood-weighted standard deviation. 
A more detailed description of the code itself can be found in \cite{Burgarella2005}, \cite{Noll2009}, and \cite{Boquien2019}.

To briefly summarize, CIGALE is designed to model a galaxy SED  by conserving the energy balance between the dust-absorbed stellar emission and its re-emission in the IR. The same strategy is also used for the AGN emission absorbed and re-emitted by the dusty torus. Many authors already presented the mechanism of the SED fitting with CIGALE including \cite{Fritz2006} AGN module \citep[i.e.][]{Buat2015, Ciesla2015, Malek2018, Wang2020, Toba2020} which assumes a central engine surrounded by a smooth dusty torus. 
In the present work SED fitting is performed using more advanced AGN module, SKIRTOR \citep{Stalevski2012,Stalevski2016}. 
SKIRTOR was implemented to the \texttt{v2020.0} version of CIGALE and tested by \citet{yang2020MNRAS_xcigale} based on the already known AGNs from the  COSMOS and the AKARI NEP field. 
This AGN module  adopts a clumpy two-phase dusty torus which can be manifested as an obstruction of the UV/optical emission from the disk.  

The SED fitting was based on a delayed star formation history with a possible additional burst, \cite{BC2003} single stellar population, and \cite{CF2000} attenuation law.
The same set of modules was used e.g. for infrared detected galaxies from the  \textit{Herschel} Extragalactic Legacy Project described in \cite{Malek2018}.

The dust emissison was calculated using  the  \cite{Dale2014} templates.  These templates are based on the sample of nearby star-forming galaxies originally presented in \cite{DaleHelou2002} with an improved modelling of emission from Polycyclic Aromatic Hydrocarbons (PAH). 
The \cite{Dale2014} model uses only one free parameter, which is the $\alpha$ power law slope in the dust mass ($M_{dust}$) and radiation field intensity ($U$) relation: $ d M_{dust} \propto U^{-\alpha} d U$. 
It simplifies the SED fitting procedure reducing the number of parameters difficult to constrain without far infrared measurements.
This module also allows one to use an optional AGN component. In the present work a dedicated AGN module was used instead.

In case of the AKARI-NEP SED fitting the SKIRTOR AGN model has been used. The adopted parameters used for the SED fitting are presented in Tab.~\ref{tab:cigale_par}. 
SEDs were fitted to objects with respect to spectroscopic redshift (training sample) and photometric redshift (training sample and output catalog of AGN candidates). The photometric redshift were taken from~\citet{ho20hsc_nep}, where it was estimated on the panchromatic NEP data using the  $LePHARE$\footnote{ \url{http://www.cfht.hawaii.edu/~arnouts/lephare.html}} software~\citep{arnouts1999MNRAS_LePHARE, ilbert2006AA_LePHARE, arnouts2011_softLePhare}.

In particular estimations of two parameters were investigated during analysis of ML output: AGN fraction ($fracAGN$), which is the AGN contribution to $L_{IR}$ in $8\mkern2mu{-}\mkern2mu1000$~$\mu$m range and AGN viewing angle ($\theta$), where $\theta = 30^{\circ}$ defines type-I and $\theta = 70^{\circ}$ corresponds to type-II.

\begin{table}[ht] 
\centering
\begin{tabular}{l|l}
\hline\multicolumn{2}{c}{star formation history:} 
\\\hline\hline
$\tau_{main}$(Myr) &  500, 2000., 4000., 7000, 9500 \\
$\tau_{burst}$(Myr) & 10000 \\
f burst &  0.0, 0.05, 0.15, 0.2 \\
age main (Myr) & 100, 500, 1800, 3500, 5000, \\
 &  6500, 8000, 9500, 10500 \\
age burst (Myr) & 100\\
\hline\multicolumn{2}{c}{single stellar population \cite{BC2003}:} \\\hline\hline
IMF  & \cite{Chabrier2003}\\
Metallicity &  0.02 \\
\hline\multicolumn{2}{c}{dust attenuation \cite{CF2000}:} \\\hline\hline
$A_{V}^{BC}$ (mag) & 0, 0.05, 0.1, 0.3, 0.5, 0.8, 1.2, 1.6, 2.0, 2.7\\
slope ISM & -0.7\\
slope BC & -0.7\\
$A_{V}^{ISM}/ A_{V}^{BC}$ & 0.8 \\
\hline\multicolumn{2}{c}{dust emission  
\cite{Dale2014}:} \\\hline\hline
fracAGN & 0\\
$\alpha$ &   0.5, 1.5, 2.0, 2.5, 3.0, 3.25, 3.75\\
\hline\multicolumn{2}{c}{AGN emission \cite{Stalevski2016}:} \\\hline\hline
 $\tau_{9.7}$ & 3, 7, 11\\
radial density & 1.0\\
angular density & 1.0\\
opening angle (deg.) & 40 \\
$r_{max}/r_{min}$ & 20\\
viewing angle ($\theta$)  & 30, 70\\
$fracAGN$ &  0, 0.1, 0.15, 0.2, 0.35, 0.55, 0.75, 0.9 \\ \hline

\end{tabular}
\caption{Modules and core input parameters used in the SED fitting with CIGALE.}
\label{tab:cigale_par}
\end{table}

\subsection{Machine learning pipeline}
\label{sec:pipeline}

\begin{figure*}[t]
\begin{subfigure}{.5\textwidth}
\centering
\includegraphics[width=0.8\textwidth]{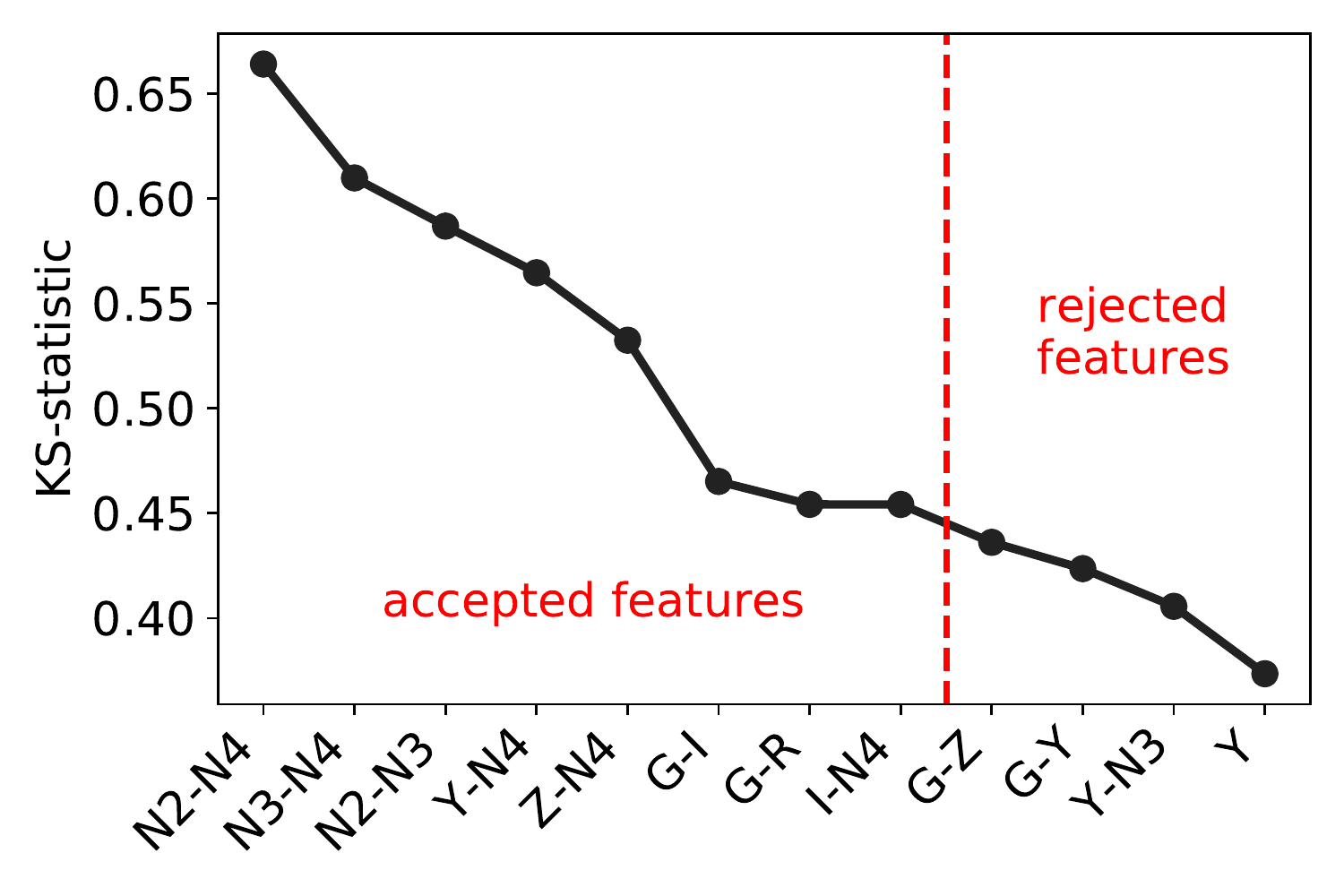}
\caption{First (main) iteration}
\label{fig:feature_selection_1st_loop}
\end{subfigure}
\begin{subfigure}{.5\textwidth}
\centering
\includegraphics[width=0.8\textwidth]{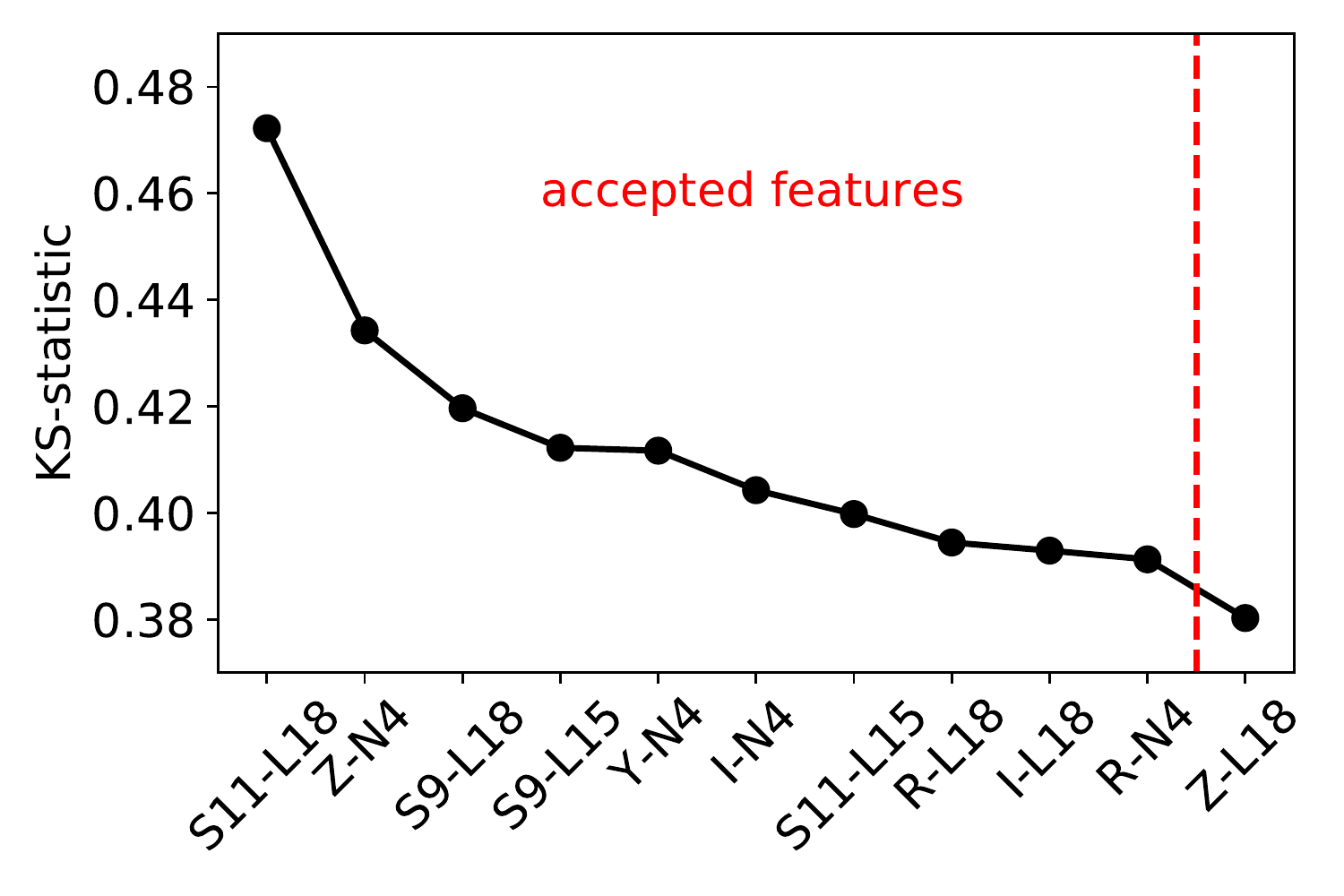}
\caption{Second iteration experiment}
\label{fig:feature_selection_2nd_loop}
\end{subfigure}
\caption{Features selected using KS-statistic value. For the first (main) iteration the KS-statistic was calculated on AGN and galaxy training samples using HSC $g$, $r$, $i$, $z$, $Y$ and IRC $N2$, $N3$, $N4$ bands and all possible colors. The optimal feature set was defined as a set where all available filters were used and the KS-statistic value was not substantially lower for less important features. For the second iteration experiment, additional MIR features in form of $S7$, $S9$, $S11$, $L15$ and $L18$ bands and all possible colors were used. In order to minimize the risk of data sparsity the selection of optimal feature set was restricted to features with the highest KS-statistic value. Only the most important features are shown.}
\label{fig:feature_selection}
\end{figure*}

The ML pipeline for AGN selection, shown in Figure~\ref{fig:pipeline}, is constructed as follows. The feature selection based on the KS-statistic was performed on the labeled data, giving the final form of a training sample. Training data, in form of flux measurements present on the selected features, was then used to determine limits of the final generalization sample.

During the training stage models described in Section~\ref{sec:algorithms_method} were tested in several settings: non-balanced and class-balanced models were applied in classical and instance-weighted (in form of error-based and distance-based fuzzy memberships) setups. First, the hyperparameter tuning was performed. Logistic regression, SVM and XGBoost models were tuned using a randomized grid searches consisting of 1000 different hyperparameters combinations. In order to search for the optimal combination, the F1 metric was maximized during the grid search. The value of the F1 metric was estimated using 100 shuffled train-test splits. An analogous split was used to calculate other evaluation metrics and their uncertainties. Finally, the selected hyperparameter combination was used to make predictions on the training data using 5-fold cross validation.

After the training, the best model was used to select the AGN candidates from the generalization sample. In order to better understand the outcome of the generalization, as well as track possible model bias uncaptured by the previous analysis, additional SED fitting tests were made on the output catalog, as well as the training set.

\section{Results} \label{sec:results}

\subsection{Feature selection}
\label{sec:feature_selection_results}

\begin{figure*}[t]
\begin{subfigure}{.48\textwidth}
\centering
\includegraphics[width=0.8\textwidth]{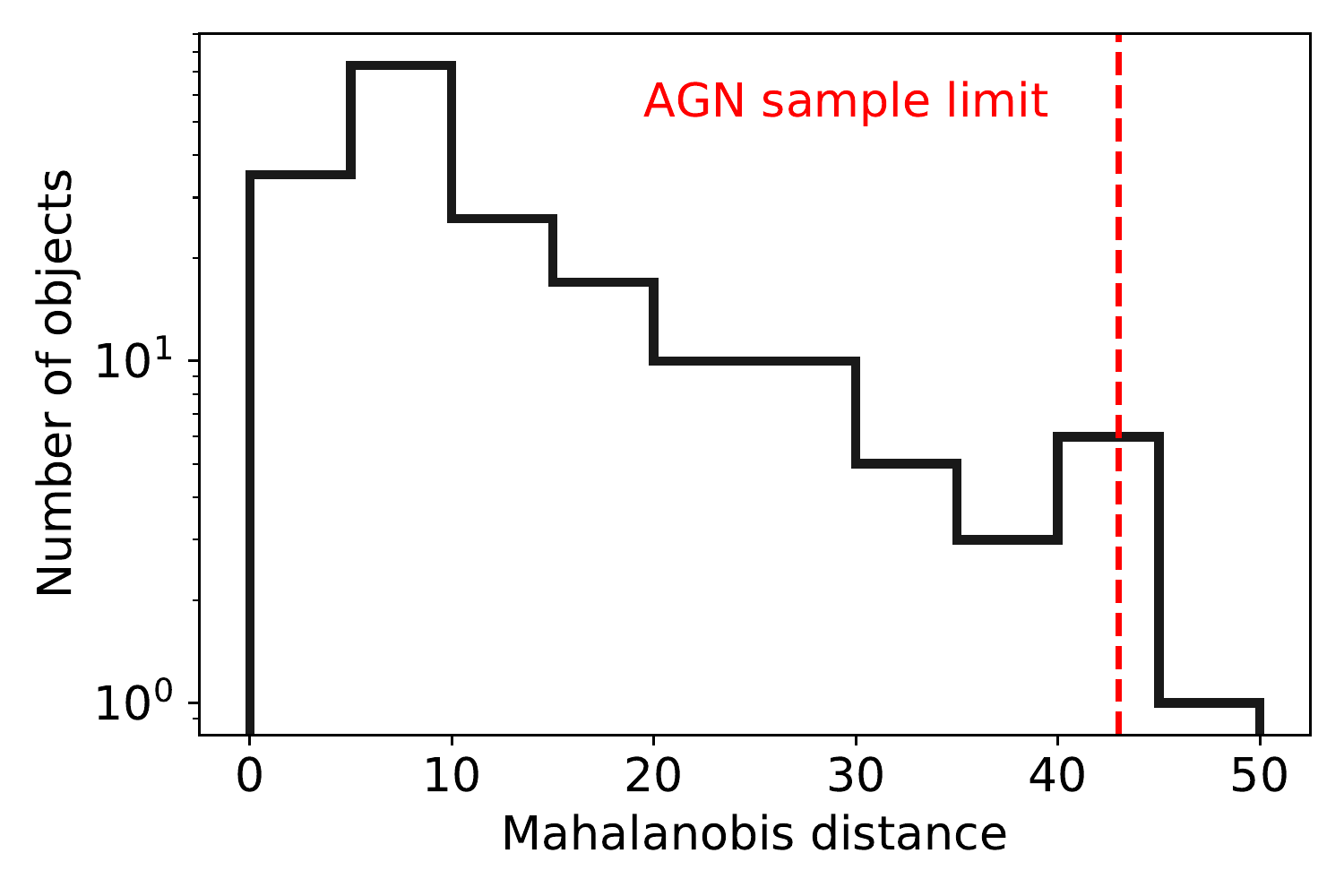}
\caption{AGN sample}
\label{fig:agn_mcd}
\end{subfigure}
\begin{subfigure}{.48\textwidth}
\centering
\includegraphics[width=0.8\textwidth]{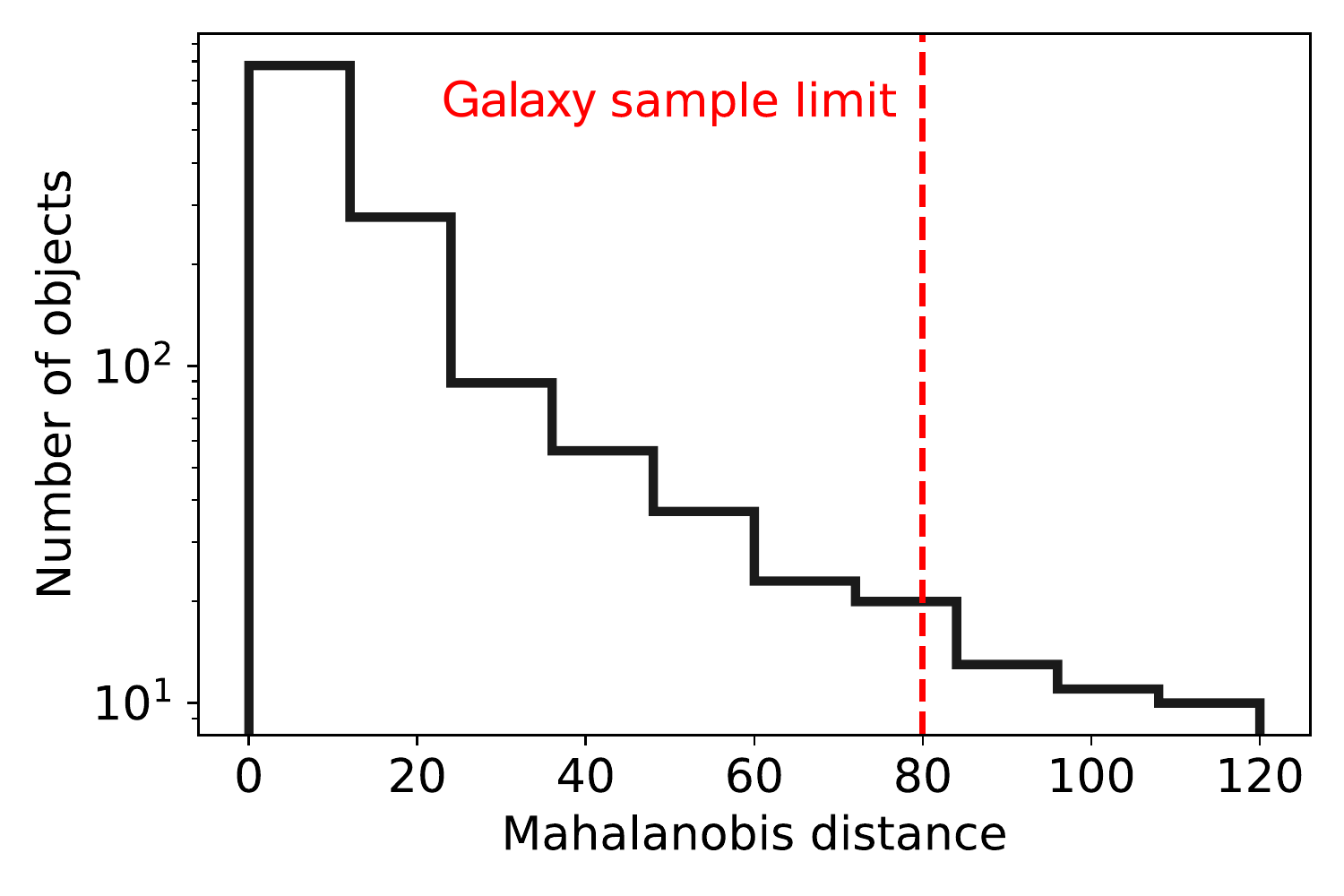}
\caption{Galaxy sample}
\label{fig:nonagn_mcd}
\end{subfigure}
\caption{Mahalanobis distance histograms for AGN and galaxy training samples. Dashed red lines correspond to a particular contamination parameter value of the MCD algorithm used to limit generalization data set to the training sample range.}
\label{fig:mcd}
\end{figure*}

In order to select features suitable for AGN extraction, the KS-statistic was calculated on the AGN and galaxy training samples using HSC $g$, $r$, $i$, $z$, $Y$ and IRC $N2$, $N3$, $N4$ bands and all possible colors. Part of the KS-statistic values for best features are shown in Fig.~\ref{fig:feature_selection_1st_loop}. The optimal feature set was defined as a set where all available filters were used and the KS-statistic value was not substantially lower for less important features. This way eight features with the highest KS-statistic values were selected as a final data representation. The selected features demonstrate an evident importance of the NIR colors as compared to VIS-based measurements.

\subsection{Generalization sample limit}
\label{sec:generalization_sample_limit_results}

The process of limiting the generalization sample to the training sample range was performed in a space constructed from the flux measurements in passbands used for the feature selection. The passband measurements were used instead of selected features in order to simplify the impact of selection effects on the generalization sample. To find the generalization sample limit, first, two different ellipsoids were fitted to the AGN and galaxy classes from the training data. Objects from the unlabeled data set, which belong to at least one of these ellipsoids were selected to a final generalization sample. The split into two separate ellipsoids was dictated by the structure of the MCD algorithm and the imbalance of the class sizes in the training data. The single ellipsoid fitted to the whole training set would treat objects from the smaller AGN class placed outside the main part of the galaxy distribution as outliers and thus reject the most important part of the feature space that allows AGN classification.

In order to find a proper contamination parameter $\alpha$ value, which defines the size of the ellipsoid, the discontinuities in the $d_M$ histograms (shown in Fig.~\ref{fig:mcd}) were sought. In the case of the AGN sample, the strong decrease in the number of objects occurs in the range of $d_M \simeq 43$, that corresponds to $\alpha_{AGN}=0.065$. In the case of the galaxy sample, a more conservative approach was used due to the higher $d_M$ dispersion. The $d_M=80$ value was selected, what corresponds to $\alpha_{GAL}=0.05$. By applying this method, the initial generalization sample size was reduced from 45 841 to 33 119 objects.

\subsection{Performance evaluation and creation of the final catalog}
\label{sec:first_iteration_results}

\begin{figure*}
\begin{subfigure}{.48\textwidth}
\centering
\includegraphics[width=0.8\textwidth]{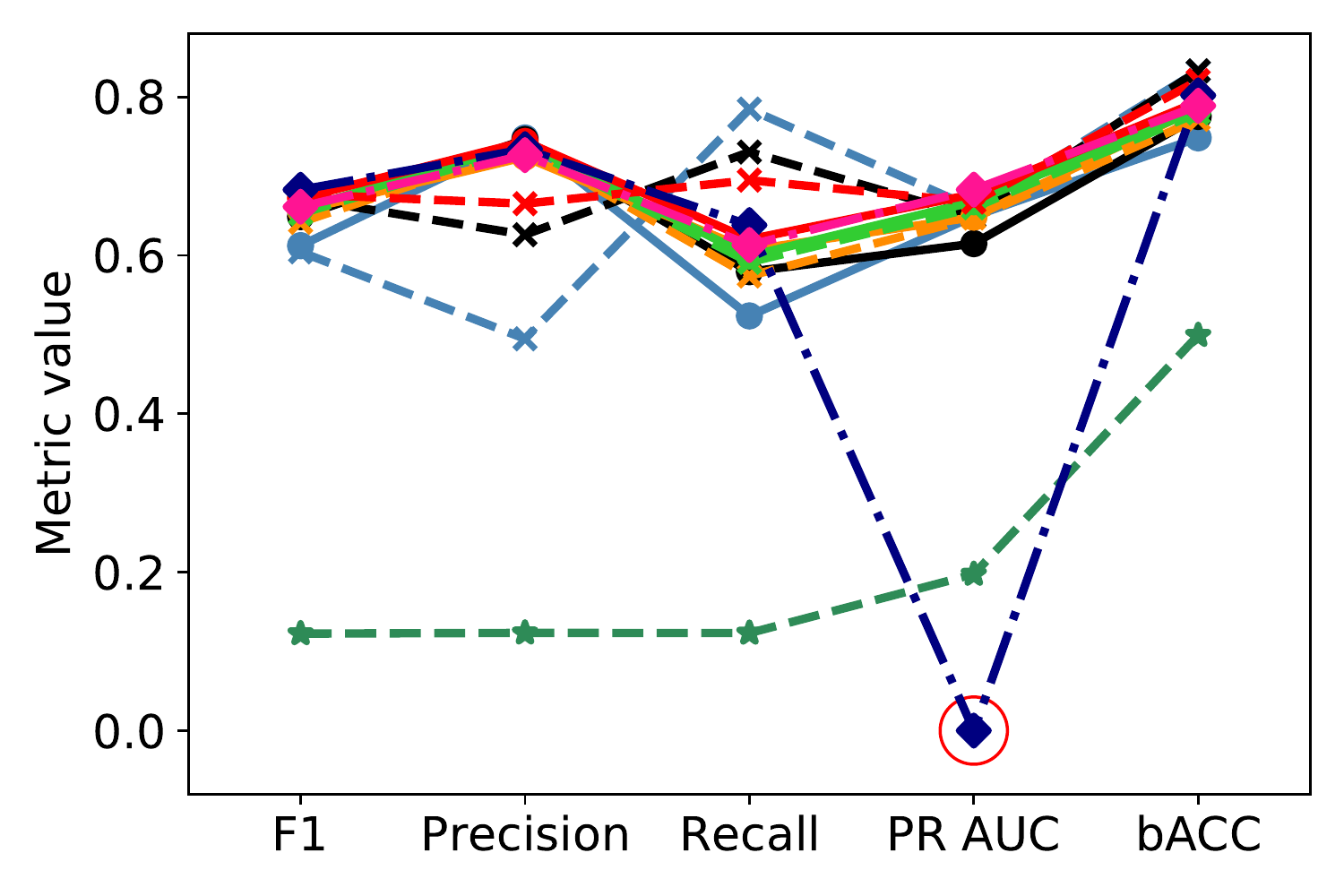}
\caption{Evaluation metric values for different classification algorithms. \\ Only models with no instance-weighting are presented.}
\label{fig:metrics_1st_iteration}
\end{subfigure}
\hspace{1em}
\begin{subfigure}{.48\textwidth}
\centering
\includegraphics[width=0.8\textwidth]{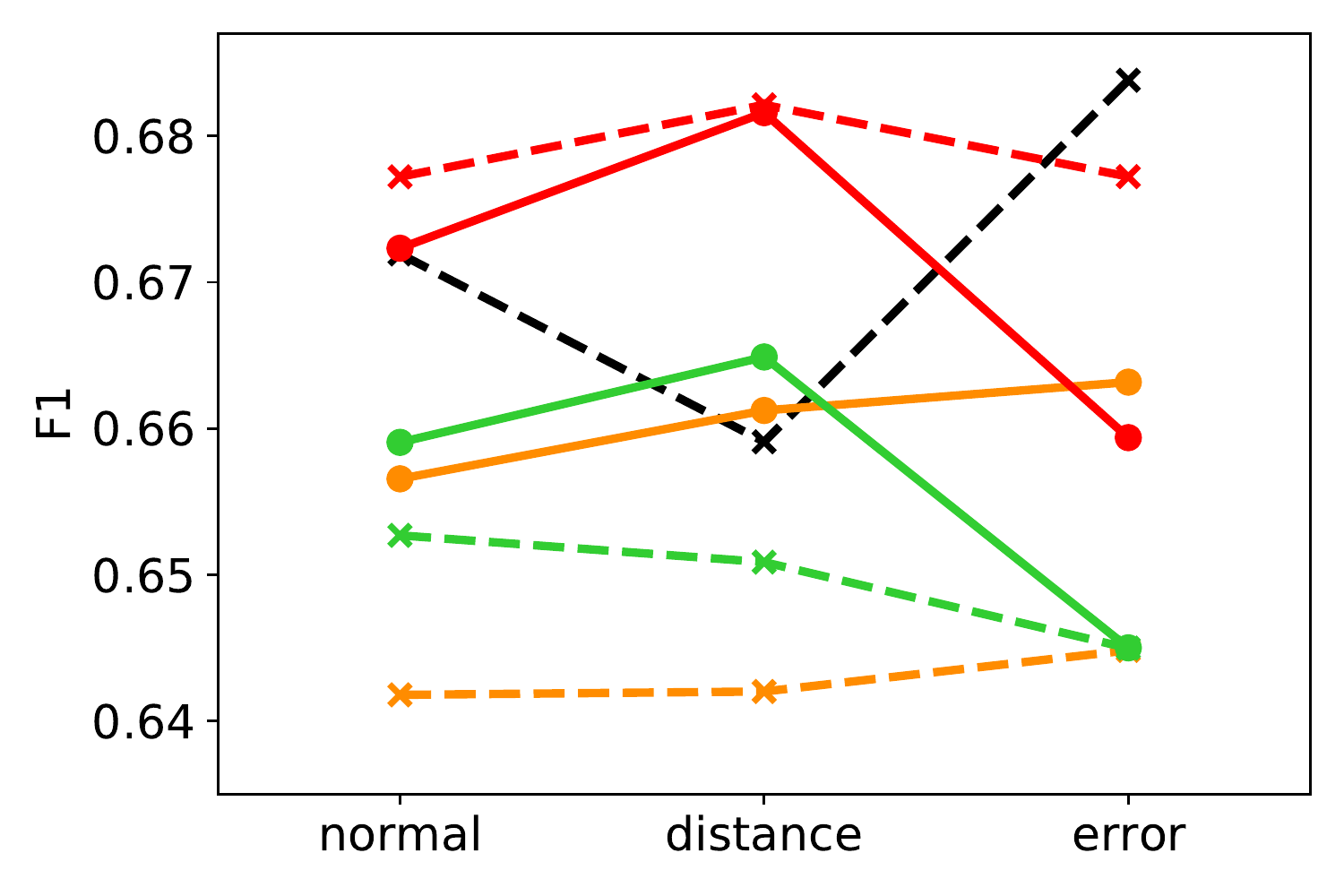}
\caption{F1 value for different instance-weighting strategies for selected types of classification algorithms.}
\label{fig:metrics_1st_F1}
\end{subfigure}

\begin{subfigure}{.48\textwidth}
\centering
\includegraphics[width=0.8\textwidth]{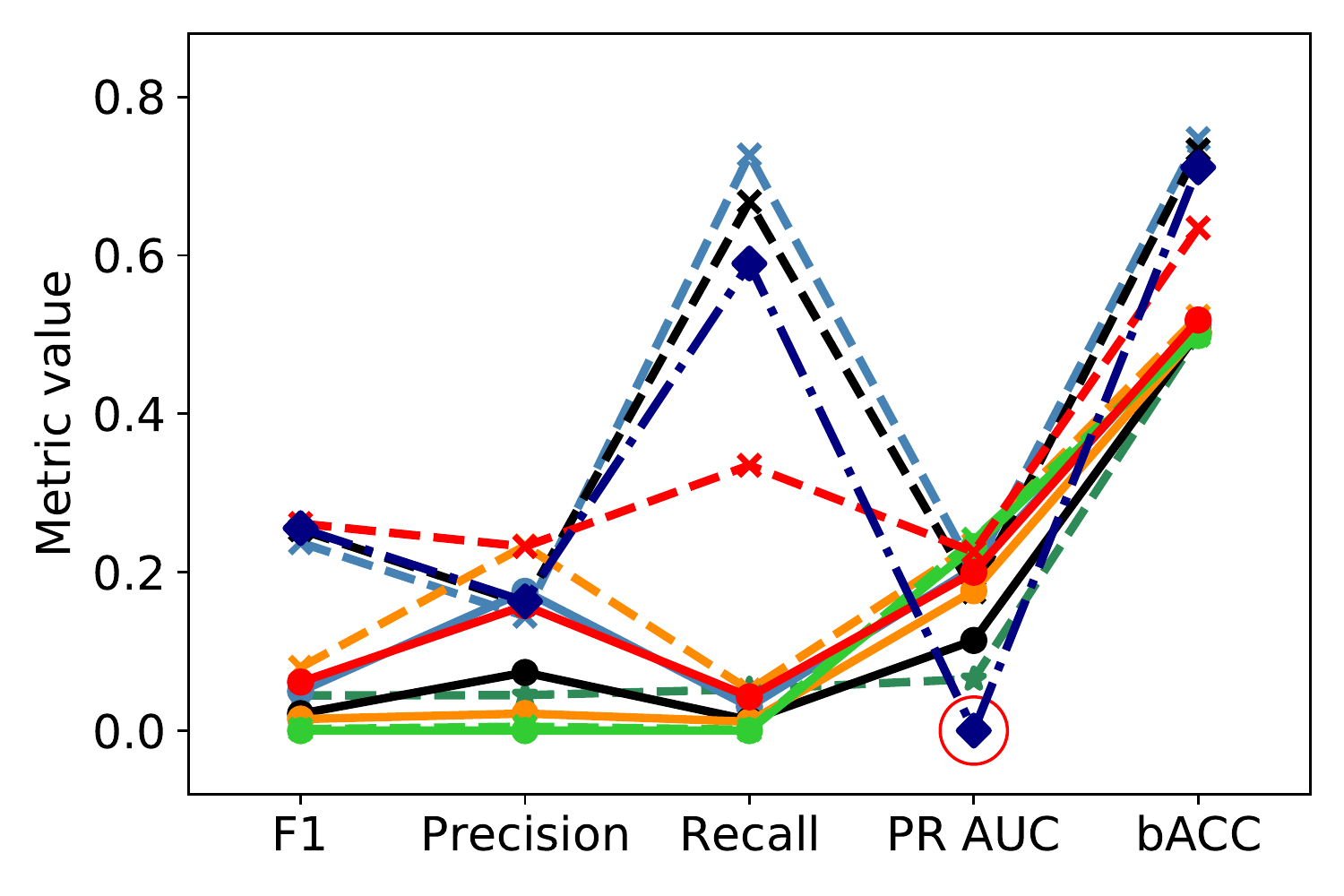}
\caption{Evaluation metric values for second iteration experiment.}
\label{fig:metrics_2nd_iteration}
\end{subfigure}
\hspace{1em}
\begin{subfigure}{.48\textwidth}
\centering
\includegraphics[width=0.8\textwidth]{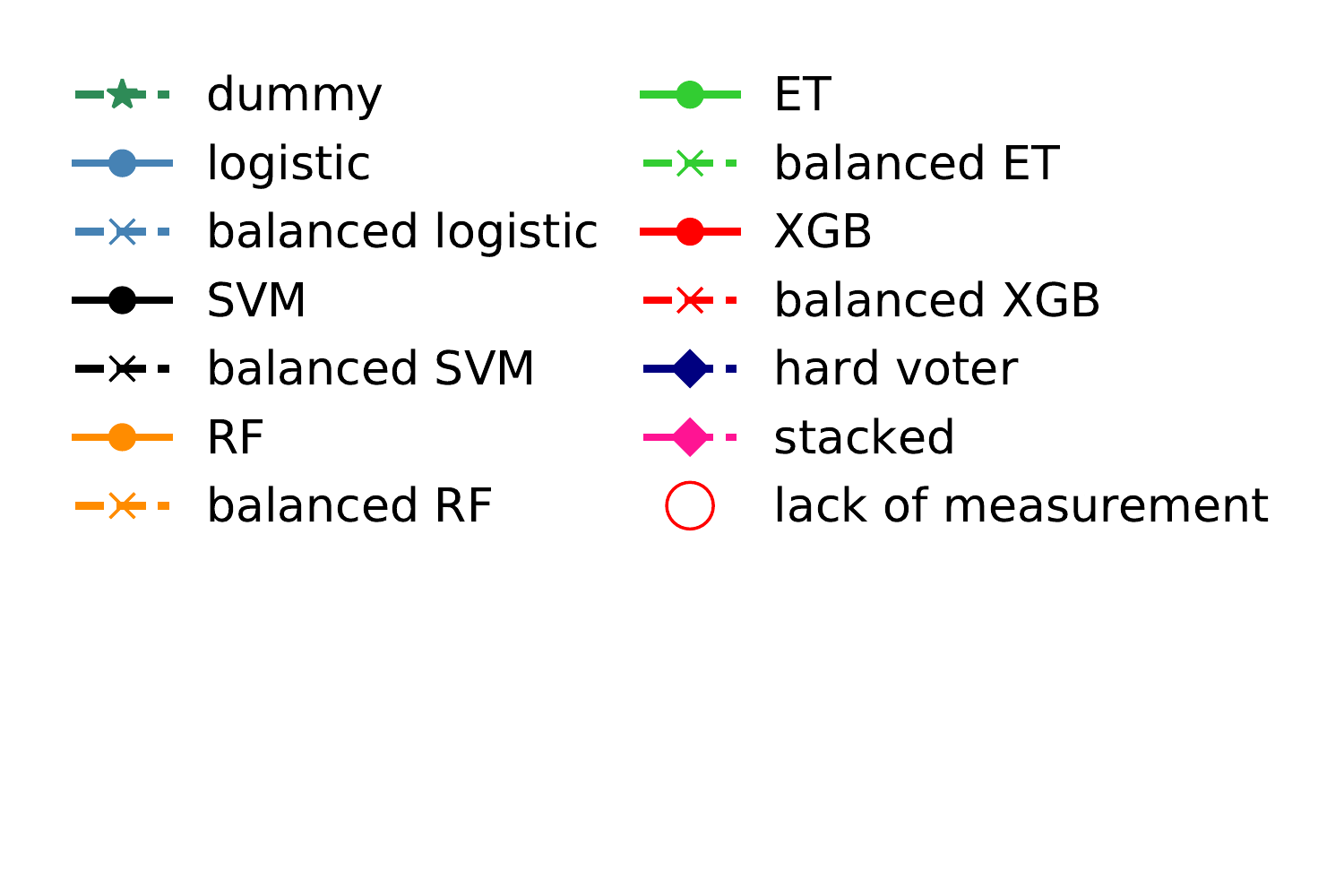}
\caption{legend}
\label{fig:metrics_legend}
\end{subfigure}

\caption{Performance evaluation of different classification models.}
\label{fig:metrics}
\end{figure*}

Each type of classification algorithm was trained and tested according to the method described in Sec.~\ref{sec:pipeline}. Detailed values of the performance evaluation can be found in Appendix in Tab.~\ref{tab:metrics_1st_iteration}. Visual comparison of performance evaluation is presented in Fig.~\ref{fig:metrics_1st_iteration}. 
In these figures one can observe a general tendency for class-balanced classifiers: an increase of recall due to the higher importance of positive class objects during training process, followed by a higher contamination of the positive candidates catalog, translated into a lower precision. During the performance comparison of non-balanced and class-balanced classifiers, several models were rejected. These were: non-balanced logistic regression due to its low recall value, class-balanced logistic regression due to its low precision and non-balanced SVM due it its low PR AUC value. The remaining classifiers showed acceptable trade-offs between precision and recall. Each one of them was used in three different instance-weight setups: uncertainty-based fuzzy membership, distance-based fuzzy membership and no instance weight. Performance comparison of these three types of models in terms of F1 metric (which was the main metric optimized during the grid search) is presented in Fig.~\ref{fig:metrics_1st_F1}. It shows not only a major impact of instance weighting on classification performance but also the opposite tendencies of error and distance weighting with respect to the non-weighted models. One can distinguish two groups of classifiers showing similar behavior: class-balanced SVM and XGBoost in all three instance-weight types preserve high recall at the expense of precision, while the rest of classifiers show the opposite tendency (see Appendix~\ref{tab:metrics_1st_iteration}). In order to benefit from both of these tendencies, all remaining classifiers were used for the construction of two types of voters: hard voter and stacked classifier. Their performance is also shown in Figure~\ref{fig:metrics_1st_iteration}. Due to the lack of probability estimation in the case of hard voter, the PR AUC measurement is missing.

A small amount of training data causes difficulties in the performance evaluation and relatively low values of both precision and recall, compelling one to focus on a classifier that can provide more certain information (higher precision) about a smaller number (lower recall) of the positive class candidates. In the present work, the best results were shown by the non-balanced fuzzy distance XGBoost classifier in the form of the most optimal precision-recall trade off (manifested also in a high F1 score). The hard voting classifier showed almost identical metric values except for a slightly lower precision (and lack of the PR AUC measurement), which is still situated within the uncertainty of XGBoost precision measurement. Since the hard voting classifier allows one to additionally reduce the variance of the final model, it was selected as the final classification scheme characterized by 0.73 precision and 0.64 recall. Generalization performed on the unlabeled data set provided a catalog of 465 AGN candidates (1.4$\%$ of the generalization sample).

\begin{figure*}[t]
\begin{subfigure}{.48\textwidth}
\centering
\includegraphics[width=0.85\textwidth]{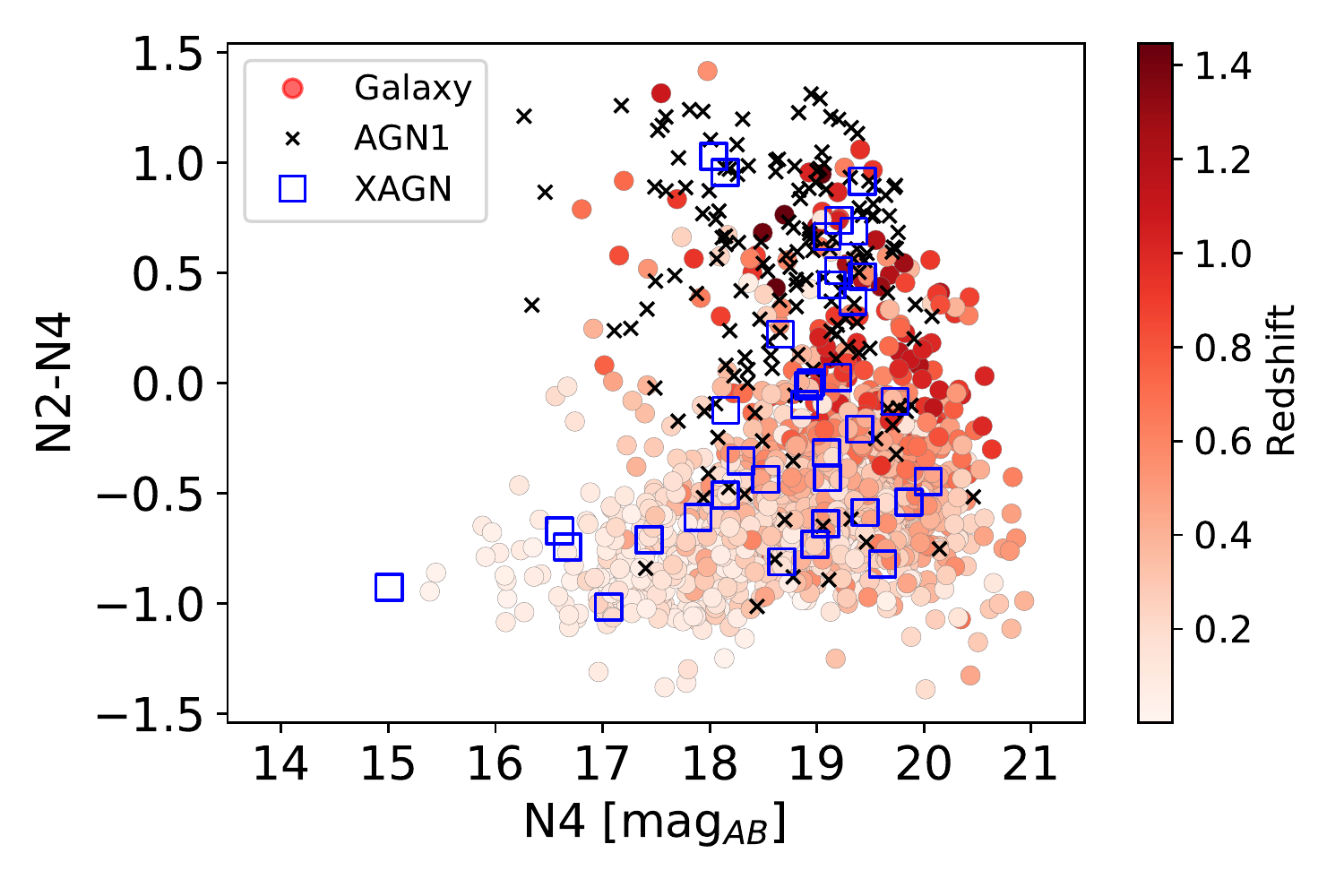}
\caption{Properties of the training data. Color of galaxies correspond to spectroscopic redshift value.}
\label{fig:n4_n2n4_train}
\end{subfigure}
\hspace{1em}
\begin{subfigure}{.48\textwidth}
\centering
\includegraphics[width=0.85\textwidth]{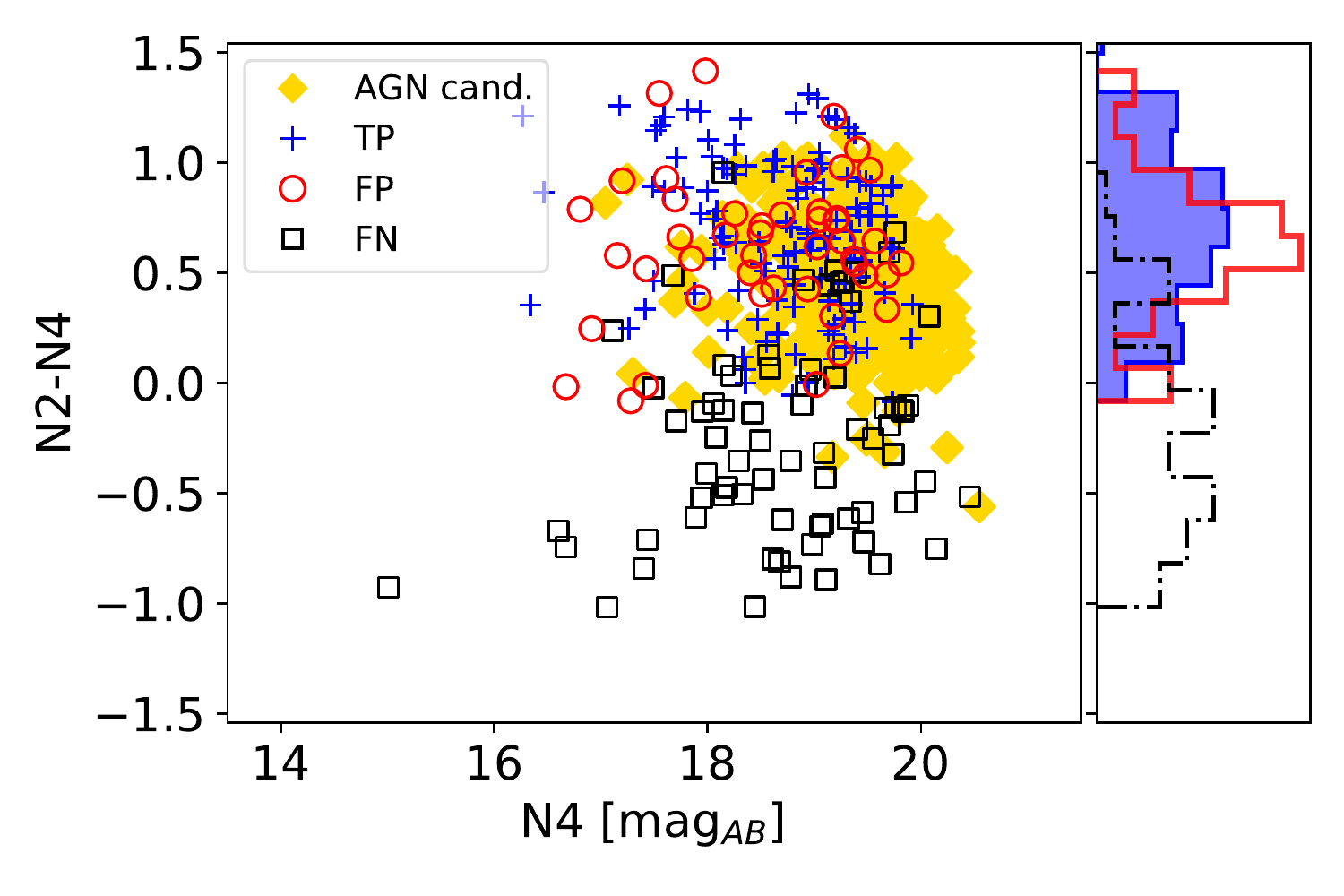}
\caption{Fist iteration training and final AGN catalog properties. Right part of the figure shows density histogram for components of the labeled set with corresponding color.}
\label{fig:n4_n2n4_1st}
\end{subfigure}

\begin{subfigure}{.48\textwidth}
\centering
\includegraphics[width=0.85\textwidth]{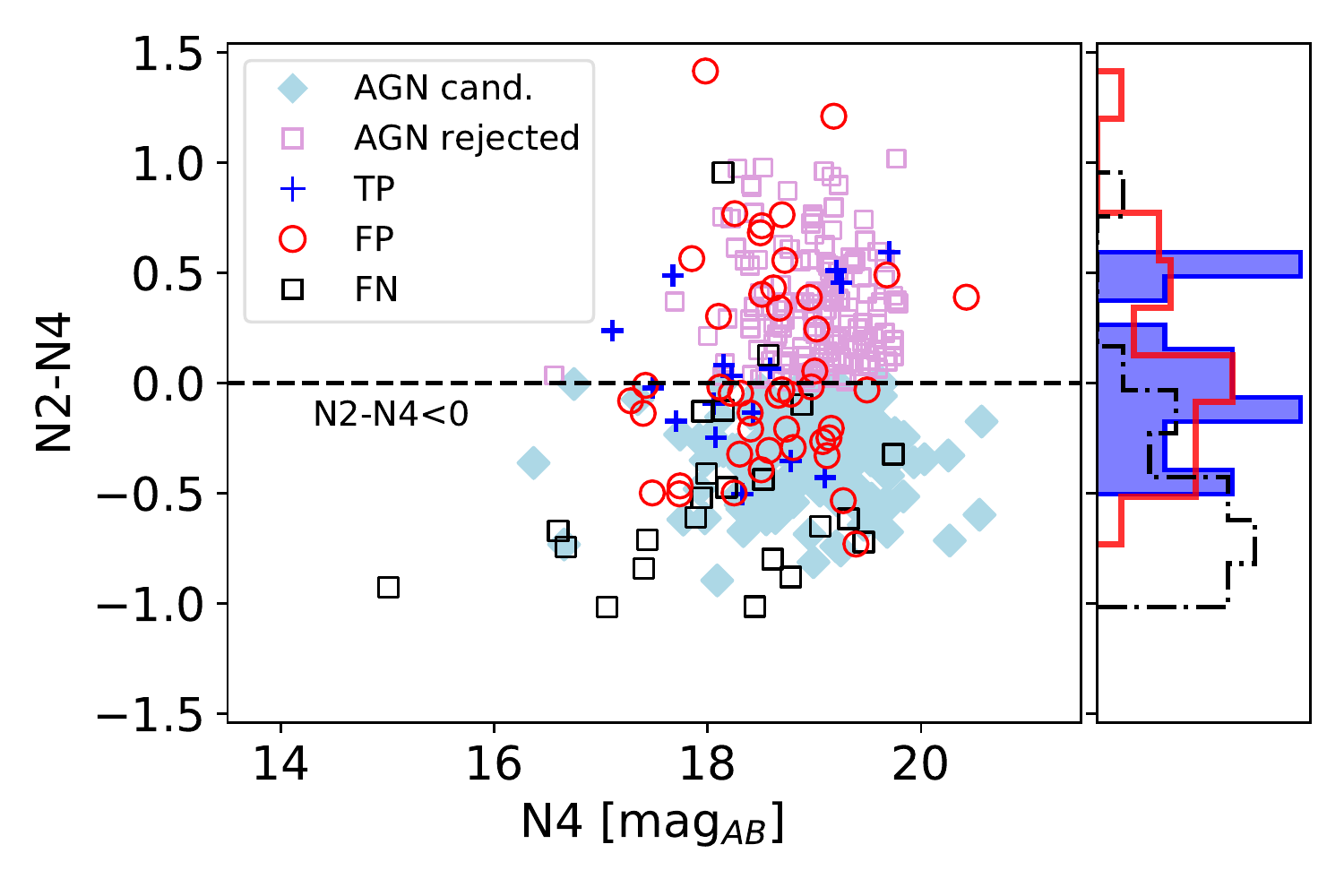}
\caption{Second iteration experiment. Modified training sample is shown as well as two parts of generalization: "rejected" objects, which occupy the red $N2\mkern2mu{-}\mkern2muN4$ color range and new objects below this range. These candidates were not included in final catalog. Right part of the figure shows density histogram for components of the labeled set with corresponding color.}
\label{fig:n4_n2n4_2nd}
\end{subfigure}
\hspace{1em}
\begin{subfigure}{.48\textwidth}
\centering
\includegraphics[width=0.85\textwidth]{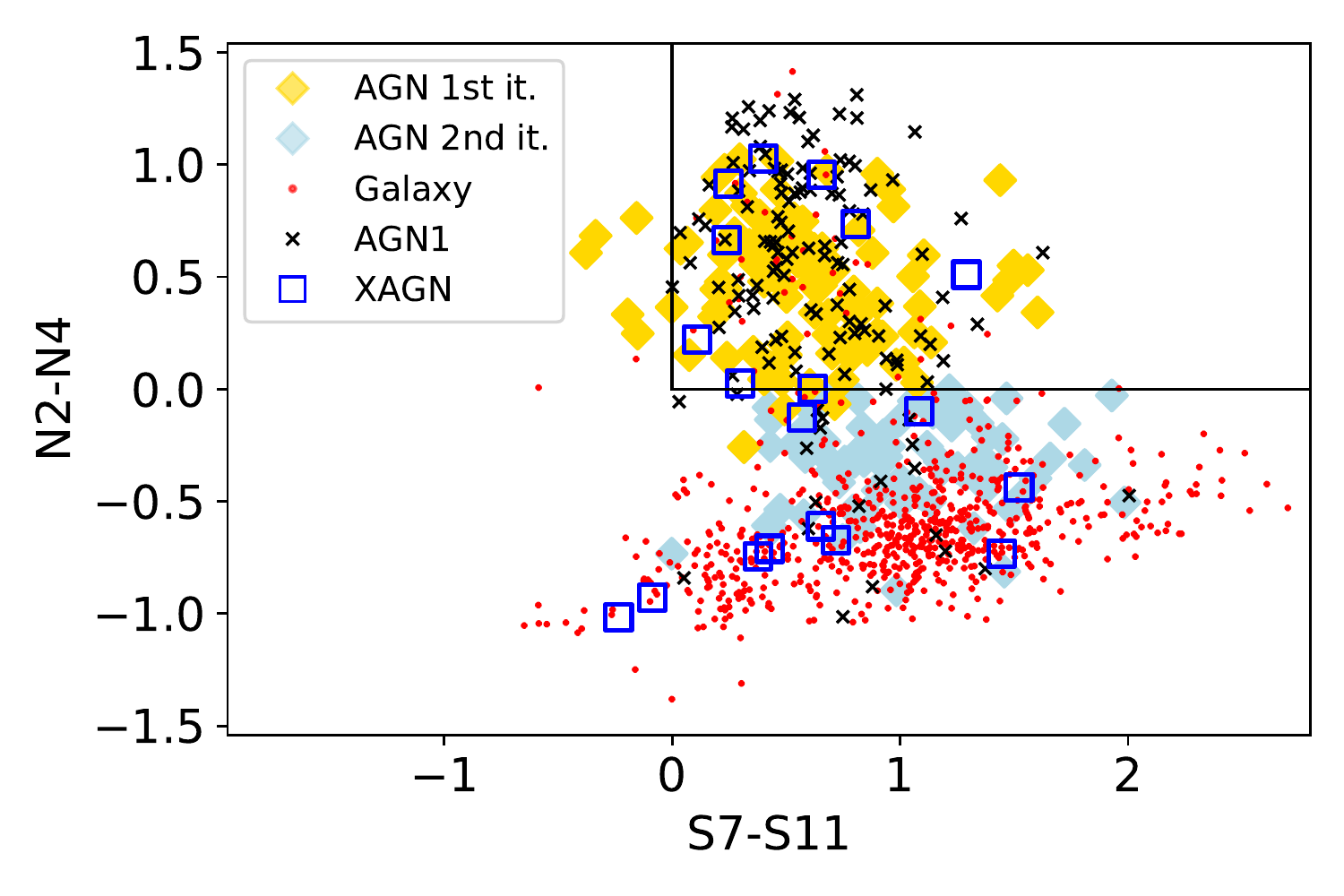}
\caption{Color-color plot used for \citealp{lee07} AGN selection. Its selection criteria demarcate the upper right square, marked by the black lines. Training objects and AGN candidates selected during the first and the second iteration from the present work, are shown in form of different markers.}
\label{fig:s7s11_n2n4}
\end{subfigure}

\caption{NIR and MIR properties of the training set and classification results. Predictions for the labeled data shown in~\ref{fig:n4_n2n4_1st}, \ref{fig:n4_n2n4_2nd} and \ref{fig:s7s11_n2n4} are compound classifications of the test data from 5-fold cross validation.}
\label{fig:N3_N2N4_both}
\end{figure*}

The  AGN candidates CM plot of $N4$ vs $N2\mkern2mu{-}\mkern2muN4$ is shown in Figure~\ref{fig:n4_n2n4_1st}. The output catalog recreates general properties of a preselected AGN target sample from~\citet{shim13nepspec}: it has a high number of proper detections around the AGN class center recovering most of type-I AGN with the main contamination source (FP) in form of high-z SFG. Another characteristic property is the impossibility of efficient class separation in the galaxies-dominated region showing mainly false negative (FN) results in the $N2\mkern2mu{-}\mkern2mu N4 \mkern2mu{<}\mkern2mu 0$ area. During analysis of training objects classification presented in CM and color-color plots, one should keep in mind that predicted labels were assigned during 5-fold CV process and might show an underestimated performance (compared to the generalization) in some feature space regions due to an absence of essential objects in the training sample.

\begin{table}[]
    \centering
    \begin{tabular}{lllll}
    
    \textbf{band} &  \textbf{median} & \textbf{MAD} & \textbf{min.} & \textbf{max.} \\
    \hline 
    g  &  22.613 &  1.139 &  18.768 &  27.586 \\
    r  &  22.061 &  0.963 &  18.764 &  25.544 \\
    i  &  21.569 &  0.815 &  18.532 &  24.347 \\
    z  &  21.292 &  0.725 &  18.165 &  23.708 \\
    Y &  21.137 &  0.674 &  18.283 &  23.430 \\
    N2 &  19.974 &  0.415 &  17.341 &  20.846 \\
    N3 &  19.687 &  0.416 &  17.289 &  20.648 \\
    N4 &  19.515 &  0.405 &  17.041 &  20.546 \\

    \hline
    \end{tabular}
    \caption{Statistical properties of the final catalog of AGN candidates. Median, median absolute deviation (MAD) and minimal and maximal limits are shown.}
    \label{tab:finalcat_stat}
\end{table}

\subsection{Second iteration experiment}

Lack of detections in the blue part of $N2\mkern2mu{-}\mkern2muN4$ distribution shows that information contained in optical and NIR passbands is not sufficient to select AGN in that region. Classifiers tends to define most of the blue AGN objects  as galaxy representatives. In order to address this issue, a supplementary classification experiment with additional information from MIR passbands ($S7$, $S9W$, $L15$, $L18W$) was performed. The main classification described in the previous section will be referred to as the \textit{first iteration}, while the supplementary experiment will be called the \textit{second iteration}. In the second iteration the training AGN class was formed from false negative cases of the first iteration, while the galaxy class remained the same, so the training sample consisted of the 744 objects in total with 39 AGNs.
In order to create an experiment that would be easily comparable to the first iteration, the second iteration generalization sample was selected as a subset of the first iteration generalization sample. It was done by applying an additional requirement of MIR detection in previously mentioned filters. Such a sample contained 2 207 objects.

Feature selection, also based on KS-statistic, enabled to choose 10 features well suited for this specific task. 
Since some of the filters were present only in features with a low KS-statistic value, the assumption that all of the filters should be utilized was omitted.
In order to minimize the risk of data sparsity the selection of optimal feature set was restricted to features with the highest KS-statistic value. The best features are shown in Fig.~\ref{fig:feature_selection_2nd_loop}. It is worthy of note that features with the highest KS-statistic value most of the time consist of MIR measurements.

Performance evaluation is presented in detail in Appendix Tab.~\ref{tab:metrics_2nd_iteration}. It is also partially shown in Fig.~\ref{fig:metrics_2nd_iteration}. This time only class-balanced SVM, class-balanced logistic regression and class-balanced XGBoost were used for the hard voter creation. The stacked classifier was not used in this case due to the high risk of overfitting caused by the small amount of data. In the second iteration the highest precision was the main priority due to its very low value. The best results of a precision-recall trade off were shown by class-balanced fuzzy distance XGBoost of $0.25\mkern2mu{\pm}\mkern2mu0.11$ and $0.37\mkern2mu{\pm}\mkern2mu0.16$ respectively. This model was selected as the final classifier for the second iteration AGN selection. Generalization performed by the second iteration classifier gives 354 AGN candidates. They are presented on CM plot alongside classified labeled samples in Fig.~\ref{fig:n4_n2n4_2nd}. Some of them (AGN candidates are presented in form of light blue and pink squares) occupy the $N2\mkern2mu{-}\mkern2mu N4$ region avoided during the first iteration. This new behavior is caused by three factors: additional information obtained from MIR measurements allowing to work in new feature regions, modified AGN distribution in the training sample forces a change of classification boundaries and a lower precision that gives higher contamination of the final catalog. Since the second iteration was supposed to be a supplement for the main, first iteration, only objects with $N2\mkern2mu{-}\mkern2muN4<0$ carry meaningful information supplementary to the main catalog. This cut limits an additional AGN catalog from 354 to 198 objects (9$\%$ of the generalization sample). Due to very high contamination and low completeness, classification performed in the second iteration is treated only as a test ground to determine limits of both the training sample and the information content of HSC and IRC data. Second iteration generalization is not included in the final AGN candidates catalog. Statistical properties of the final catalog obtained during first iteration are presented in Tab.~\ref{tab:finalcat_stat}.

\subsection{Comparison with color cut method}

\begin{figure*}[t]
\begin{subfigure}{.5\textwidth}
\centering
\includegraphics[width=0.85\textwidth]{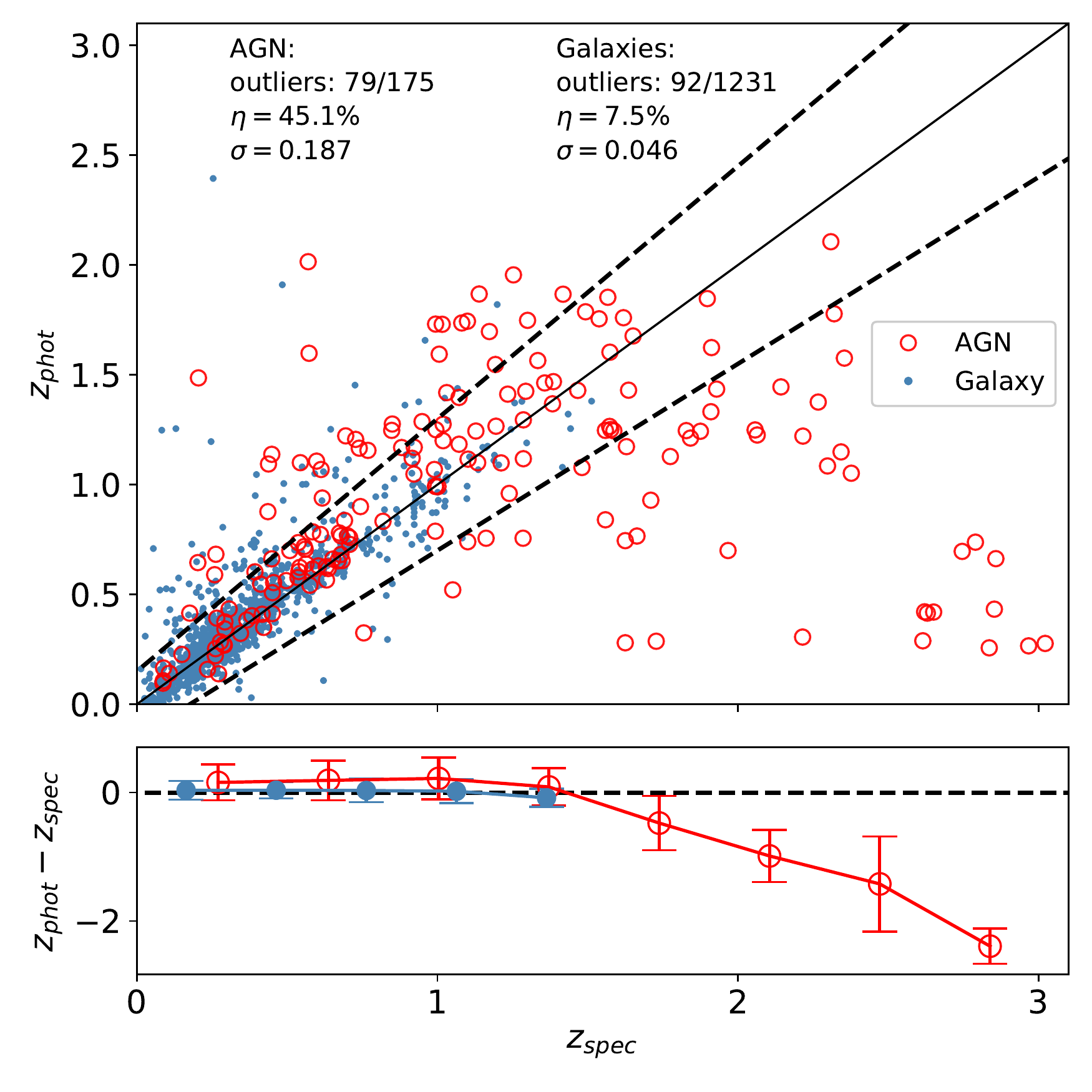}
\caption{Comparison between spectroscopic redshift and photometric redshift estimation~\citep{ho20hsc_nep} for training galaxy and AGN samples. The dotted lines refer to $z_{phot}=z_{spec}\pm 0.15(1+z_{spec})$ cone. The $\eta$ describes the fraction of outliers defined as objects outside the cone. $\sigma$ is the normalised median absolute deviation $1.48\times median(|\Delta z|/(1+z)$. Lower plot shows the mean residuals with standard deviations. Only $z<3$ objects are shown.}
\label{fig:redshift_comparison}
\end{subfigure}
\hspace{1em}
\begin{subfigure}{.5\textwidth}
\centering
\includegraphics[width=0.85\textwidth]{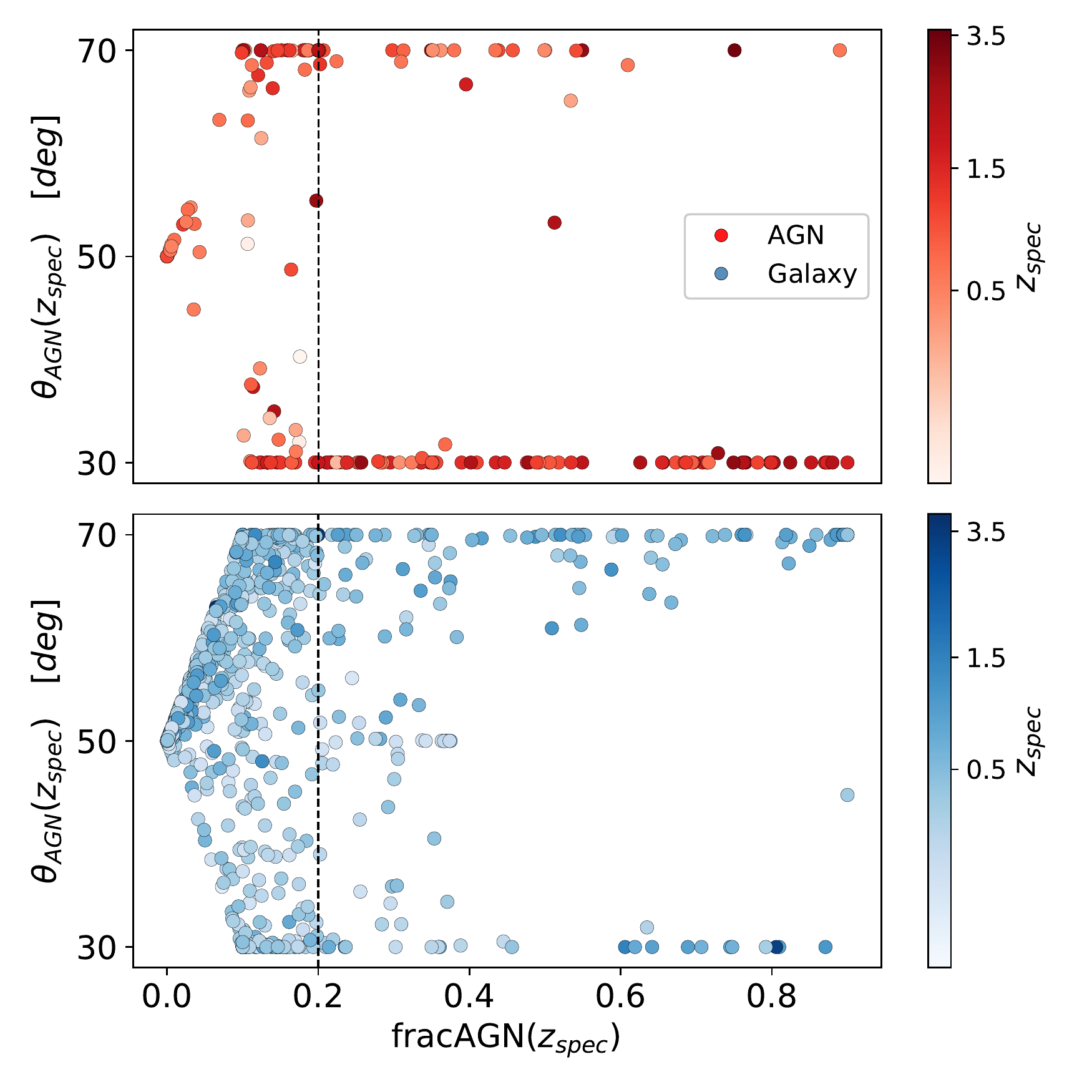}
\caption{Dependence of an AGN viewing angle ($\theta$) on AGN fraction ($fracAGN$) for AGN and galaxy training samples. Both parameters were estimated by SED fitting using spectroscopic redshifts. $\theta=30^{\circ}$ corresponds to pure type-I AGN and  $\theta=70^{\circ}$ to pure type-II with intermediate types in between. Color intensity corresponds to the spectroscopic redshift value. Dotted line shows $fracAGN=0.2$ - value which defines AGN$_{SED}$.}
\label{fig:frac_vs_theta}
\end{subfigure}
\caption{Visualization of parameters used during SED analysis on the training data. XAGN were excluded from this analysis due to lack of spectroscopic redshift.}
\label{fig:sed_fitting}
\end{figure*}

Since the training sample used in the present work is mainly based on NIR+MIR target preselection described in~\citet{lee07} it is crucial to investigate how ML-based selection performs with respect to this color cut method. In the original NIR+MIR method, AGN selection was performed on a $S11\mkern2mu{<}\mkern2mu18.5$~mag limited sample in order to increase completeness of the output catalog. Comparison of the two methods presented in this paper was performed without this condition, since the $S11$ limit is impossible to apply in optical+NIR combined space. Fig.~\ref{fig:s7s11_n2n4} shows a $N2\mkern2mu{-}\mkern2muN4$ vs $S7\mkern2mu{-}\mkern2muS11$ color-color plot. The color cut method reflects the power-law properties of the AGN SED, which should cause red MIR colors. The upper right square ($N2\mkern2mu{-}\mkern2muN4\mkern2mu{>}\mkern2mu0$, $S7\mkern2mu{-}\mkern2muS11\mkern2mu{>}\mkern2mu0$) used to select AGN in the \cite{lee07} method is also a location of the majority of AGN candidates (yellow points). Objects selected during the second iteration are located in between the AGN and galaxy classes giving an indication that part of these objects might be low activity AGN. It is also worth notice that a large fraction XAGN are located far outside AGN class center escaping both selection methods.

Tab.~\ref{tab:lee_comparison} shows performance comparison between the color-cut method and the classification scheme formulated in the present work. Hard voter shows higher precision and balanced accuracy values, but lower F1 and recall. Such values are consistent with precision-recall trade-off established during model selection, when precision was recognized as a more important metric. Both the color-color plot and  the metric values show good consistency between the two methods. However, in order to perform such comparison both the training sample and output catalog were limited to objects detected in $S7$ and $S11$ passbands. The training sample was reduced from 1 547 to 815 objects and the final catalog from 465 to 113 objects. Despite similar metric values, the color-cut method was able to detect only 24\% of objects selected in the present work.

\begin{table}[]
    \centering
    \begin{tabular}{lllll}
    
    \textbf{Method} &  \textbf{F1} & \textbf{Precision} & \textbf{Recall} & \textbf{bACC} \\
    \hline 
    $\begin{cases}
      N2-N4>0 \\
      S7-S11>0 \\
    \end{cases}$ &  0.76 &  0.73 &  0.80 &  0.84 \\
    hard voter  &  0.75 &  0.77 &  0.74 &  0.86
    \\

    \hline
    \end{tabular}
    \caption{Performance comparison of final selected  classifier with color-cut method described in~\citealp{lee07}. Only objects detected in $S7$ and $S11$ passbands were used.}
    \label{tab:lee_comparison}
\end{table}

\subsection{SED fitting analysis} \label{sec:results_sed}

Fig.~\ref{fig:redshift_comparison} shows a comparison between spectroscopic and photometric redshifts used for the SED fitting analysis. For this analysis the training sample was used. XAGNs, whose redshifts have not been measured, were excluded from further analysis. Good agreement characterize both AGN and galaxy classes up to redshift 1.5. The higher redshift range, occupied almost exclusively by AGN, shows strong bias towards $z_{phot}$ underestimation. Fig.~\ref{fig:frac_vs_theta} shows a viewing angle dependency on AGN fraction calculated using spectroscopic redshifts. Several observations crucial for catalog analysis can be made. 

As it was already observed in~\citet{Ciesla2015}, constraining the AGN contribution to the total infrared luminosity is challenging for objects with a small $fracAGN$ value. In order to perform a reliable analysis, only objects with $fracAGN\mkern2mu{>}\mkern2mu0.2$ were used. A similar $fracAGN$ limit was applied in ~\citet{Ciesla2015, Malek2018}. These objects will be referred to as SED-identified AGN (AGN$_{SED}$).

The SED-based AGN type identification is done based on AGN properties in UV, MIR and FIR. The type-I AGN is showing strong emission in UV and MIR. The type-II AGN, devoid of strong UV emission, is characterized by large FIR to MIR emission ratio~\citep{Ciesla2015}. Due to the small number of objects with detections in FIR, distinction between the two types might be disturbed. Moreover, as it was pointed out in \citet{mountrichas2020arXiv_cigale_type_bias}, lack of X-ray measurements creates a tendency towards type-I assignment. The majority of AGN$_{SED}$ in the AGN training sample were identified as type-I AGN. The galaxy sample shows the opposite tendency towards type-II representation, which might be even higher due to the present bias caused by lack of X-ray measurements. Since the $fracAGN$ definition is based on IR properties of the SED, this plot is a strong indication of the presence of a different, narrow line AGN class in the NEP data. Most of galaxies from training sample classified as type-I AGN showed strong AGN UV and MIR contribution. These might be AGN characterized by low S/N ratio or weak emission lines in the optical range.

More precise information is presented in Tab.~\ref{tab:sed_results} that shows the properties of AGN$_{SED}$ objects within the training sample and the final AGN catalog. It shows an approximately 50\% discrepancy between spectroscopic and SED AGN identification within the AGN training sample in all redshift ranges. The difference between the two methods is even higher considering $\sim13\%$ of AGN$_{SED}$ are detected in galaxy sample. Completeness  and purity loss in the low-z range after application of photometric redshifts manifests itself in a decrease of type-I-dominated AGN$_{SED}$. Photometric redshift systematics above $z=1.5$ results in a substantial completeness loss. The catalog of AGN candidates shows intermediate properties between AGN and galaxy training samples at low photometric redshift. In the high redshift range the training AGN sample and AGN candidates have similar SED AGN contribution and a smaller number of type-I SED AGN in the case of final catalog.

The SED-fitting analysis in the present work is not very extensive due to several reasons. Firstly, as it was already pointed out, AGNs selected according to their emission line properties are, at least partially, a different class of objects than galaxies with high AGN fraction recognized through the SED fitting (only 50\% of spectroscopically-defined AGNs have also high AGN fraction fitted). This behaviour is not caused by the low quality of the SED-fitting for AGN from the training set as the distributions of $\chi^2$ for spectroscopic AGNs and other objects with $fracAGN>0.2$ are very similar. The final ML-based classifier was trained to reproduce the MIR-based selection, which, in turn, was constructed to reflect the structure of the spectroscopically selected AGN training sample, which consists mostly of type I AGNs. Thus studying properties of spectroscopic-type AGN class misclassifications through SED-defined AGN class may easily lead to a false conclusions.
Secondly, the data in both training sample and the final AGN catalog are highly inhomogeneous in terms of photometric coverage. Because of that their SEDs cannot be analysed as a whole in a statistically consistent manner. To avoid this problem one would have to clean the catalog, which would result in a strong decrease in the data volume. Such in-depth study of SEDs was presented in \citet{Wang2020}. However, due to the different constraints on the data properties, a sample analyzed in~\citet{Wang2020} has only a few common objects with both training data and resultant catalog discussed in the present paper. This discrepancy makes further comparison almost impossible.

Future works may include more extensive AGN classification based on training sets making use of SED templates or both SED templates and spectral features, which should allow not only to identify AGN candidates, but also to make classifiers sensitive to chosen types of AGNs. In the present work the main limitation is the spectroscopy based training sample, making the classifier sensitive primarily to type I AGNs.

\begin{table*}[ht] 
\centering
\begin{tabular}{lccccccccc}
\textbf{Sample} & \textbf{Size} & \textbf{F$_{specz}$} & $\Delta$\textbf{frac$_{specz}$} & \textbf{F}$_{specz}^{typeI}$ & \textbf{C loss} & \textbf{P loss} & \textbf{F}$_{photz}$ & $\Delta$\textbf{frac}$_{photz}$ & \textbf{F}$_{photz}^{typeI}$ \\
\hline 
\multicolumn{10}{c}{\textbf{$z_{phot}\leq1.5$}} \\
\hline
AGN training & 160 & 53\% & 6\% & 72\% & 14\% & 11\% & 50\% & 5\% & 55\%\\
Galaxy training & 1249 & 13\% & 3\% & 29\% & 3\% & 6\% & 16\% & 3\% & 20\% \\
AGN candidates & 293 & $-$ &  $-$ & $-$ & $-$ & $-$ & 25\% & 6\% & 42\% \\
\hline 
\multicolumn{10}{c}{\textbf{$z_{phot}>1.5$}} \\
\hline
AGN training & 27 & 44.0\% & 6\% & 83\% & 22\% & 11\% & 33\% & 6\% & 55\% \\
AGN candidates & 160 & $-$ & $-$ & $-$ & $-$ & $-$ & 38\% & 6\% & 27\% \\
\hline

\end{tabular}
\caption{SED fitting results calculated for training samples and obtained AGN candidates catalog in two photometric redshift $z_{phot}$ ranges (below and above $z_{phot}=1.5$). $F$ shows percentage of SED-identified AGN in sample (defined as $fracAGN\mkern2mu{>}\mkern2mu0.2$) obtained using spectroscopic ($z_{spec}$) and photometric ($z_{phot}$) redshift respectively. $F^{typeI}$ defines what fraction of SED-identified AGN (on the ground of spectroscopic or photometric redshifts) exhibit dominance of type-I AGN features ($\theta<50^{\circ}$). The $\Delta$\textit{frac} is a median of AGN fraction parameter estimation uncertainty based on $z_{spec}$ and $z_{phot}$ respectively. Completeness loss (\textit{C loss}) is the percentage of SED-identified AGN objects which lost AGN characteristic after switching from $z_{spec}$ to $z_{phot}$. Purity loss (\textit{P loss}) is the percentage of objects falsely identified as AGN after switching from $z_{spec}$ to $z_{phot}$. The \textit{Size} column corresponds to the number of objects in the sample.}

\label{tab:sed_results}
\end{table*}

\section{Summary} \label{sec:summary}

In the present work several important results were achieved. From the perspective of the main aim of the project, it was possible to successfully reproduce \citet{lee07} NIR+MIR AGN classification only with optical and NIR data. Both methods showed consistent results in NIR+MIR space, but only 24\% of objects in the final AGN candidate catalog was detected in MIR. Hence, the new method shows substantial improvement in the NEP AGN detection, by preserving the effectiveness of the method in a much larger volume of available data. To obtain satisfying results, a large number of different model types were tested. It was shown that voting schemes, built on an ensemble of models, might improve classification results even when a small amount of training data are available. Additional tests of instance weighting showed that physically or geometrically motivated weights have a major impact on the classification performance and thus further research in this field should be done. 

The second iteration experiment pushed classification towards galactic locus - typically occupied by a large XAGN fraction. The result was a drastic increase of catalog contamination.
Efficient AGN selection in this region is impossible with training data constructed from currently available spectroscopic and X-ray detections. SED fitting analysis of training data showed a large amount of AGN which are not identified with spectroscopic observations. SED fitting analysis of the output catalog of AGN candidates shows intermediate properties between the AGN and galaxy training samples in terms of the $fracAGN$ and $\theta$. This might be caused by both contamination from high-z SFG or presence of AGN types that are not identified by SED fitting. There is a significant difference between SED-identified and spectroscopically identified AGNs - both selection methods trace different AGN classes. Thus, it is impossible to directly compare the performance of ML-based (trained on spectroscopical class labels) and SED-fitting based methods. This result shows also a possible way to overcome difficulties with AGN selection outside the AGN class center: training sample augmentation with template-based objects might give a classifier crucial information  about problematic regions of feature space and extend selection to other types of AGN.

\begin{acknowledgements}
Authors are very thankful to the referee  for  the  comments  and  corrections  provided  to  the manuscript.  They allowed us to create much more understandable and transparent paper.
This research was conducted under the agreement on scientific cooperation between the Polish Academy of Sciences and the Ministry of Science and Technology in Taipei and supported by the Polish National Science Centre grant UMO-2018/30/M/ST9/00757 and by Polish Ministry of Science and Higher Education grant DIR/WK/2018/12.
K.M. is supported by the Polish National Science Centre grant UMO-2018/30/E/ST9/00082. A.D. is supported by the Polish National Science Centre grant UMO-2015/17/D/ST9/02121. 
T.G. acknowledges the support by the Ministry of Science and Technology of Taiwan through grant108-2628-M-007-004-MY3. 
H.S.H. was supported by the New Faculty Startup Fund from Seoul National University.
\end{acknowledgements}

\bibliographystyle{aa} 
\bibliography{akari_nep_wide.bib} 

\onecolumn
\appendix \label{sec:appendix}
\appendixpage

\section{Software} 
\label{sec:software}

For the ML pipeline construction and performance analysis SciPy~\citep{virtanen2020SciPy}, NumPy~\citep{harris2020NumPy}, Pandas~\citep{2010pandas, reback2020pandas}, Scikit-learn~\citep{scikit-learn} and XGBoost~\citep{chen16} python packages were used. Results visualization was done with Matplotlib~\citep{matplotlib} and Seaborn~\citep{seaborn} packages. SED fitting was performed via CIGALE~\citep{Boquien2019} software.

\section{Metric values} 
\label{sec:appendix_metrics}

This appendix presents details of the performance evaluation. Metric values presented in the Tab.~\ref{tab:metrics_1st_iteration} refer to the main training (first iteration). The best classifier from this set was used for the creation of the final AGN candidates catalog. Tab.~\ref{tab:metrics_2nd_iteration} shows results from the second iteration experiment.

\begin{table*}[]
    \centering
    \begin{tabular}{llllll}
        \textbf{Classifier} &        \textbf{F1} &  \textbf{Precision} &    \textbf{Recall} &    \textbf{PR AUC} &      \textbf{bACC} \\ 
        \hline \\
        dummy classifier &  0.12 &   0.12 &  0.12 &  0.20 &  0.50 \\
        \hline \\
        \multicolumn{6}{c}{\textbf{Logistic regression:}} \\
        \noalign{\vskip 1mm } 
        \hline
        non-balanced normal &  0.61$\pm$0.06 &  0.75$\pm$0.09 &  0.52$\pm$0.07 &  0.65$\pm$0.08 &  0.75$\pm$0.04 \\         
        class-balanced normal &  0.60$\pm$0.05 &  0.49$\pm$0.06 &  0.78$\pm$0.07 &  0.66$\pm$0.07 &  0.83$\pm$0.03 \\ \\      
        
        non-balanced fuzzy error &  0.61$\pm$0.05 &  0.75$\pm$0.07 &  0.52$\pm$0.07 &  0.66$\pm$0.07 &  0.75$\pm$0.03 \\   
        class-balanced fuzzy error &  0.59$\pm$0.05 &  0.48$\pm$0.06 &  0.78$\pm$0.07 &  0.64$\pm$0.08 &  0.83$\pm$0.03 \\ \\
        
        non-balanced fuzzy distance &  0.64$\pm$0.05 &  0.72$\pm$0.07 &  0.57$\pm$0.06 &   0.65$\pm$0.07 &   0.77$\pm$0.03 \\ 
        class-balanced fuzzy distance &    0.60$\pm$0.06 &  0.49$\pm$0.06 &  0.78$\pm$0.07 &   0.65$\pm$0.08 &  0.83$\pm$0.03 \\
        
        \hline \\
        \multicolumn{6}{c}{\textbf{SVM:}} \\
        \noalign{\vskip 1mm } 
        \hline
        
        non-balanced normal &  0.65$\pm$0.06 &  0.75$\pm$0.06 &  0.58$\pm$0.08 &  0.61$\pm$0.08 &  0.77$\pm$0.04 \\ 
        class-balanced normal &  0.67$\pm$0.05 &  0.63$\pm$0.07 &  0.73$\pm$0.06 &  0.65$\pm$0.07 &  0.83$\pm$0.03 \\ \\
        
        non-balanced fuzzy error &  0.63$\pm$0.06 &  0.74$\pm$0.07 &  0.56$\pm$0.08 &   0.61$\pm$0.08 &  0.76$\pm$0.04 \\    
        class-balanced fuzzy error &   0.68$\pm$0.05 &  0.64$\pm$0.06 &  0.74$\pm$0.07 &  0.65$\pm$0.08 &   0.84$\pm$0.03 \\ \\ 
        
        non-balanced fuzzy distance &  0.67$\pm$0.05 &  0.75$\pm$0.07 &  0.60$\pm$0.07 &   0.62$\pm$0.08 &  0.79$\pm$0.03 \\
        class-balanced fuzzy distance &  0.66$\pm$0.05 &  0.60$\pm$0.06 &  0.73$\pm$0.05 &  0.64$\pm$0.07 &  0.83$\pm$0.03 \\
        
        \hline \\
        \multicolumn{6}{c}{\textbf{Random forest:}} \\
        \noalign{\vskip 1mm } 
        \hline
        non-balanced normal &  0.66$\pm$0.06 &   0.72$\pm$0.08 &  0.61$\pm$0.07 &  0.65$\pm$0.09 &  0.79$\pm$0.03 \\
        class-balanced normal &  0.64$\pm$0.06 &  0.74$\pm$0.07 &  0.57$\pm$0.08 &  0.65$\pm$0.08 &  0.77$\pm$0.04 \\ \\
        
        non-balanced fuzzy error &  0.66$\pm$0.05 &  0.72$\pm$0.06 &  0.62$\pm$0.07 &  0.65$\pm$0.07 &  0.79$\pm$0.03 \\
        class-balanced fuzzy error &  0.64$\pm$0.06 &  0.74$\pm$0.08 &  0.57$\pm$0.07 &  0.66$\pm$0.08 &  0.77$\pm$0.04 \\ \\
        
        non-balanced fuzzy distance &  0.66$\pm$0.05 &  0.73$\pm$0.07 &  0.61$\pm$0.07 &   0.65$\pm$0.08 &  0.79$\pm$0.03 \\
        class-balanced fuzzy distance &  0.64$\pm$0.06 &  0.74$\pm$0.09 &  0.57$\pm$0.06 &  0.65$\pm$0.08 &   0.77$\pm$0.03 \\
        
        \hline \\
        \multicolumn{6}{c}{\textbf{Extremely randomized trees:}} \\
        \noalign{\vskip 1mm } 
        \hline
        
        non-balanced normal &  0.66$\pm$0.05 &  0.74$\pm$0.07 &  0.60$\pm$0.07 &  0.67$\pm$0.07 &  0.78$\pm$0.03 \\
        class-balanced normal &  0.65$\pm$0.06 &   0.74$\pm$0.07 &  0.59$\pm$0.08 &   0.66$\pm$0.08 &   0.78$\pm$0.04 \\  \\    
        
        non-balanced fuzzy error &  0.64$\pm$0.07 &  0.73$\pm$0.08 &  0.59$\pm$0.08 &  0.66$\pm$0.08 &  0.78$\pm$0.04 \\
        class-balanced fuzzy error &   0.64$\pm$0.06 &   0.73$\pm$0.07 &  0.58$\pm$0.07 &  0.65$\pm$0.08 &  0.78$\pm$0.04 \\  \\
        
        non-balanced fuzzy distance &  0.66$\pm$0.06 &  0.75$\pm$0.07 &  0.60$\pm$0.07 &  0.66$\pm$0.08 &  0.79$\pm$0.04 \\ 
        class-balanced fuzzy distance &  0.65$\pm$0.06 &  0.73$\pm$0.08 &   0.59$\pm$0.07 &   0.65$\pm$0.08 &   0.78$\pm$0.03 \\
        
        \hline \\
        \multicolumn{6}{c}{\textbf{XGBoost:}} \\
        \noalign{\vskip 1mm } 
        \hline
        
        non-balanced normal &  0.67$\pm$0.06 &  0.74$\pm$0.07 &   0.62$\pm$0.08 &  0.68$\pm$0.08 &  0.79$\pm$0.04 \\         
        class-balanced normal &  0.68$\pm$0.06 &  0.66$\pm$0.08 &  0.69$\pm$0.06 &   0.67$\pm$0.08 &  0.82$\pm$0.03 \\      \\
        
        non-balanced fuzzy error &  0.66$\pm$0.06 &  0.74$\pm$0.08 &  0.60$\pm$0.06 &  0.67$\pm$0.07 &  0.78$\pm$0.03 \\    
        class-balanced fuzzy error &   0.68$\pm$0.06 &  0.66$\pm$0.07 &  0.70$\pm$0.08 &  0.67$\pm$0.08 &  0.82$\pm$0.04 \\ \\   
        
        non-balanced fuzzy distance &  0.68$\pm$0.06 &  0.74$\pm$0.07 &   0.64$\pm$0.08 &  0.68$\pm$0.08 &   0.80$\pm$0.04 \\ 
        class-balanced fuzzy distance &  0.68$\pm$0.05 &   0.65$\pm$0.07 &  0.72$\pm$0.06 &  0.66$\pm$0.08 &  0.83$\pm$0.03 \\
        
        \hline \\
        \multicolumn{6}{c}{\textbf{Voting schemes:}} \\
        \noalign{\vskip 1mm } 
        \hline
        
        stacked classifier &  0.66$\pm$0.05 &  0.73$\pm$0.08 &  0.61$\pm$0.07 &  0.68$\pm$0.08 &  0.79$\pm$0.03 \\
        hard voter &  0.68 &   0.73 &  0.64 &       \textemdash &  0.80 \\
        \hline
        
    \end{tabular}
    \caption{Metrics for the main training (first iteration).}
    \label{tab:metrics_1st_iteration}
\end{table*}

\begin{table*}[]
    \centering
    \begin{tabular}{llllll}
    \hline
    
    \textbf{Classifier} &        \textbf{F1} &  \textbf{Precision} &    \textbf{Recall} &    \textbf{PR AUC} &      \textbf{bACC} \\ 
    \hline \\
    dummy classifier &  0.04 &   0.04 &  0.05 &  0.06 &  0.50 \\
    \hline \\
    \multicolumn{6}{c}{\textbf{Logistic regression:}} \\
    \noalign{\vskip 1mm } 
    \hline
    
    non-balanced normal &   0.05$\pm$0.09 &  0.18$\pm$0.36 &   0.03$\pm$0.06 &  0.20$\pm$0.11 &  0.51$\pm$0.03 \\         
    class-balanced normal &   0.24$\pm$0.07 &  0.14$\pm$0.05 &  0.73$\pm$0.18 &    0.20$\pm$0.10 &  0.75$\pm$0.09 \\      \\
    
    non-balanced fuzzy error &  0.09$\pm$0.09 &  0.17$\pm$0.21 &  0.07$\pm$0.08 &  0.19$\pm$0.10 &   0.52$\pm$0.04 \\    
    class-balanced fuzzy error &  0.17$\pm$0.05 &  0.10$\pm$0.03 &  0.80$\pm$0.16 &  0.17$\pm$0.09 &  0.71$\pm$0.07 \\   \\
    
    non-balanced fuzzy distance &  0.07$\pm$0.12 &   0.19$\pm$0.32 &   0.05$\pm$0.08 &   0.17$\pm$0.09 &    0.52$\pm$0.04 \\ 
    class-balanced fuzzy distance &  0.27$\pm$0.09 &  0.17$\pm$0.07 &  0.68$\pm$0.19 &  0.20$\pm$0.10 &  0.74$\pm$0.09 \\

    \hline \\
    \multicolumn{6}{c}{\textbf{SVM:}} \\
    \noalign{\vskip 1mm } 
    \hline
    
    non-balanced normal &  0.02$\pm$0.06 &  0.07$\pm$0.23 &  0.01$\pm$0.04 &  0.11$\pm$0.07 &   0.50$\pm$0.02 \\         
    class-balanced normal &  0.25$\pm$0.08 &   0.16$\pm$0.06 &  0.67$\pm$0.16 &   0.17$\pm$0.09 &  0.73$\pm$0.08 \\     \\ 
    
    non-balanced fuzzy error &  0.06$\pm$0.11 &  0.11$\pm$0.19 &   0.05$\pm$0.09 &  0.14$\pm$0.08 &  0.52$\pm$0.05 \\    
    class-balanced fuzzy error &  0.20$\pm$0.08 &  0.12$\pm$0.05 &  0.55$\pm$0.18 &   0.20$\pm$0.13 &  0.67$\pm$0.09 \\   \\
    
    non-balanced fuzzy distance &  0.00$\pm$0.02 &  0.00$\pm$0.02 &  0.00$\pm$0.01 &  0.08$\pm$0.05 &  0.49$\pm$0.006 \\ 
    class-balanced fuzzy distance &   0.24$\pm$0.07 &  0.16$\pm$0.05 &  0.55$\pm$0.15 &  0.15$\pm$0.07 &  0.69$\pm$0.07 \\
    
    \hline \\
    \multicolumn{6}{c}{\textbf{Random forest:}} \\
    \noalign{\vskip 1mm } 
    \hline
    
    non-balanced normal &  0.01$\pm$0.05 &  0.02$\pm$0.08 &  0.01$\pm$0.04 &  0.18$\pm$0.09 &   0.50$\pm$0.02 \\
    class-balanced normal &    0.08$\pm$0.14 &  0.23$\pm$0.39 &  0.05$\pm$0.09 &  0.23$\pm$0.15 &  0.52$\pm$0.05 \\ \\ 
    
    non-balanced fuzzy error &  0.02$\pm$0.07 &  0.03$\pm$0.12 &  0.01$\pm$0.07 &  0.20$\pm$0.11 &  0.50$\pm$0.04 \\ 
    class-balanced fuzzy error &  0.08$\pm$0.13 &  0.23$\pm$0.36 &  0.05$\pm$0.08 &  0.24$\pm$0.12 &  0.52$\pm$0.04 \\ \\
    
    non-balanced fuzzy distance &  0.00$\pm$0.03 &    0.01$\pm$0.07 &  0.00$\pm$0.02 &    0.18$\pm$0.10 &  0.50$\pm$0.01 \\ 
    class-balanced fuzzy distance &  0.11$\pm$0.13 &   0.32$\pm$0.40 &    0.07$\pm$0.09 &  0.25$\pm$0.12 &  0.53$\pm$0.04 \\
    
    \hline \\
    \multicolumn{6}{c}{\textbf{Extremely randomized trees:}} \\
    \noalign{\vskip 1mm } 
    \hline
    
    non-balanced normal &      0.0$\pm$0.0 &     0.0$\pm$0.0 &      0.0$\pm$0.0 &  0.23$\pm$0.12 &  0.499$\pm$0.002 \\         
    class-balanced normal &  0.0$\pm$0.02 &  0.0$\pm$0.05 &  0.0$\pm$0.01 &  0.24$\pm$0.13 &  0.50$\pm$0.01 \\   \\  
    
    non-balanced fuzzy error &      0.0$\pm$0.0 &     0.0$\pm$0.0 &      0.0$\pm$0.0 &   0.24$\pm$0.11 &  0.498$\pm$0.003 \\    
    class-balanced fuzzy error &      0.0$\pm$0.0 &     0.0$\pm$0.0 &      0.0$\pm$0.0 &  0.24$\pm$0.14 &  0.499$\pm$0.002 \\  \\
    
    non-balanced fuzzy distance &      0.0$\pm$0.0 &     0.0$\pm$0.0 &      0.0$\pm$0.0 &  0.24$\pm$0.13 &  0.499$\pm$0.002 \\
    class-balanced fuzzy distance &      0.0$\pm$0.0 &     0.0$\pm$0.0 &      0.0$\pm$0.0 &  0.22$\pm$0.11 &  0.499$\pm$0.002 \\
    
    \hline \\
    \multicolumn{6}{c}{\textbf{XGBoost:}} \\
    \noalign{\vskip 1mm } 
    \hline
    
    non-balanced normal &  0.06$\pm$0.11 &  0.16$\pm$0.31 &  0.04$\pm$0.08 &    0.20$\pm$0.11 &  0.52$\pm$0.04 \\         
    class-balanced normal &  0.26$\pm$0.11 &  0.23$\pm$0.11 &  0.33$\pm$0.15 &  0.23$\pm$0.11 &  0.63$\pm$0.07 \\ \\
    
    non-balanced fuzzy error &   0.07$\pm$0.11 &  0.15$\pm$0.25 &  0.05$\pm$0.08 &  0.17$\pm$0.09 &  0.52$\pm$0.04 \\    
    class-balanced fuzzy error &  0.23$\pm$0.11 &  0.18$\pm$0.10 &  0.34$\pm$0.16 &  0.20$\pm$0.10 &  0.63$\pm$0.08 \\   \\
    
    non-balanced fuzzy distance &  0.08$\pm$0.12 &   0.26$\pm$0.41 &  0.048$\pm$0.08 &  0.26$\pm$0.12 &  0.52$\pm$0.04 \\ 
    class-balanced fuzzy distance &  0.29$\pm$0.12 &  0.25$\pm$0.11 &  0.37$\pm$0.16 &   0.24$\pm$0.11 &  0.65$\pm$0.08 \\
    
    \hline \\
    \multicolumn{6}{c}{\textbf{Voting Schemes:}} \\
    \noalign{\vskip 1mm } 
    \hline
    hard voter &  0.26 &   0.16 &  0.59 &   \textemdash &  0.71 \\
    \hline
    \end{tabular}
    \caption{Metrics for the second iteration experiment}
    \label{tab:metrics_2nd_iteration}
\end{table*}

\end{document}